\documentclass[useAMS,usenatbib,aps,floatdix,nofootinbib]{mn2e}

\usepackage{color}

\usepackage{graphics}
\usepackage{amssymb,amsmath}
\usepackage{amsmath}
\usepackage{psfrag}
\usepackage{graphicx}

\newcommand{\f}{\frac}

\newcommand{\be}{\begin{equation}}      
\newcommand{\ee}{\end{equation}}      
      
\newcommand{\bef}{\begin{figure}}      
\newcommand{\eef}{\end{figure}}      
\newcommand{\bea}{\begin{eqnarray}}    
\newcommand{\eea}{\end{eqnarray}}      
  
\def\spose#1{\hbox to 0pt{#1\hss}}
\def\ltapprox{\mathrel{\spose{\lower 3pt\hbox{$\mathchar"218$}}
\raise 2.0pt\hbox{$\mathchar"13C$}}}
\def\gtapprox{\mathrel{\spose{\lower 3pt\hbox{$\mathchar"218$}}
\raise 2.0pt\hbox{$\mathchar"13E$}}}
\def\inapprox{\mathrel{\spose{\lower 3pt\hbox{$\mathchar"218$}}
\raise 2.0pt\hbox{$\mathchar"232$}}}

\def\bse{\begin{subequations}}
\def\ese{\end{subequations}}

\def\lsim{\raise 0.4ex\hbox{$<$}\kern -0.8em\lower 0.62ex\hbox{$\sim$}} 
\def\gsim{\raise 0.4ex\hbox{$>$}\kern -0.7em\lower 0.62ex\hbox{$\sim$}}

\def\f0N{f_0^{(N)}}
\def\bec{\begin{center}}
\def\eec{\end{center}}

\title[Violent and mild relaxation of an isolated self-gravitating
  uniform and spherical cloud of particles] {Violent and mild
  relaxation of an isolated self-gravitating uniform and spherical
  cloud of particles} \author[F. Sylos Labini] {Francesco Sylos
  Labini${^{1,}}{^{2}}$\\ $^1$Centro Studi e Ricerche Enrico Fermi,
  Via Panisperna 00184 - Rome - Italy\\ $^2$Istituto dei sistemi
  complessi, Consiglio Nazionale delle Ricerche, Via dei Taurini 19,
  00185 Rome, Italy\\ }
\begin{document}

\date{\today}

\maketitle

\begin{abstract}
 The collapse of an isolated, uniform and spherical cloud of
 self-gravitating particles represents a paradigmatic example of a
 relaxation process leading to the formation of a quasi-stationary
 state in virial equilibrium.  We consider several N-body simulations
 of such a system, with the initial velocity dispersion as a free
 parameter.  We show that there is a clear difference between
 structures formed when the initial virial ratio is {  $b_0 =
   2K_0/W_0 < b_0^c \approx -1/2$} and $b_0>b_0^c$. These two sets of
 initial conditions give rise respectively to a mild and violent
 relaxation occurring in about the same time scale: however in the
 latter case the system contracts by a large factor, while in the
 former it approximately maintains its original size. Correspondingly
 the resulting quasi equilibrium state is characterized by a density
 profile decaying at large enough distances as $r^{-4}$ or with a
 sharp cut-off. {  The case } $b_0<b_0^c$ can be well described by
 the {  Lynden-Bell theory of collisionless relaxation considering
   the system confined in a box}. {  On the other hand} the relevant
 feature for $b_0>b_0^c$ is the ejection of particles and energy,
 which is not captured by such a theoretical approach: {  for this
   case} we introduce a simple physical model to explain the formation
 of the power-law density profile. This model shows that the
 behavior $n(r) \sim r^{-4}$ is the typical density profile that is
 obtained when the initial conditions are cold enough that mass and
 energy ejection occurs.  In addition, we clarify the origin of the
 critical value of the initial virial ratio $b_0^c$.
\end{abstract}



\section{Introduction}

The evolution of a system of massive particles interacting solely by
Newtonian gravity is a paradigmatic problem for astrophysics,
cosmology and statistical physics. The underlying open question
concerns the relaxation mechanism that drives the system to form
structures which seem to be in a sort of equilibrium, as for instance
different kind of astrophysical objects such as globular clusters,
galaxies, and galaxy clusters
\citep{lyndenbell,Padmanabhan_1990,binney,Saslaw00,Heggie03,aarseth_book}.
In a galaxy the two body relaxation time is of order $\tau_2 \approx
10^{17}$ years \citep{binney}, and is much longer than the age of the
universe (i.e., $\approx 10^{10}$ years): {  for this reason} these
objects are not in thermal equilibrium. {  However,} they present
common features as the luminosity profiles (see e.g.,
\cite{deVaucouleurs48,binney2}).  Much theoretical work has been
devoted to study the dynamical model to characterize such profiles and
despite the numerical simulations have shown that structures formed in
some cases are compatible with observations, the physical origin { 
  of these profiles} has not been yet clarified from a theoretical
point of view.
Namely, the problem still remains to explain {  how to form} the
shape of density profiles and of velocity distributions of stellar
structures like elliptical galaxies and globular clusters {  that
  are} generally characterized by a dense central core and a dilute
halo --- where the halo is often featured by a power-law decay of the
radial density \citep{binney,binney2}.


In cosmology one faces a different but somewhat related problem.
Since more than a decade it has been realized that a major issue about
gravitational clustering dynamics concerns the formation of the
so-called halo-structures, which are considered the primary building
blocks in terms of which the non-linear structures observed in
cosmological simulations are described \citep{sheth}.  These are
approximately spherical symmetric structures, but sometimes with
complex substructures, and with a density profile that that has almost
universal statistical features and unknown dynamical origin.  Density
profiles of dark matter halos have become one of the most challenging
issues for our understanding of cold dark matter structure
formation. Numerical simulations provide evidences of steep central
density cusps with power law slopes $\rho \sim r^{-\beta}$, with
$\beta \approx 1 $ at small scales and $\beta \approx 3 $ at large
ones \citep{navarro1,navarro2,Moore1,Moore2,
  Diemand_Moore_Stadel_2004,Reed_etal_2003,navarro,merritt2006}.  { 
  Recently \cite{graham2006} showed that in simulated dark matter
  models, at large enough scales, slopes of $\beta \approx -4$ might
  be permitted.}  Several attempts have been made for an analytical
derivation of the density profile (see, e.g.,
\cite{Bertschinger_1985,Syer_and_White_1998,
  Subramanian_Cen_Ostriker_2000,Hiotelis_2002,Manrique_etal_2002,Dekel_etal_2003}
and references therein), and none seem to present a clear and simple
explanation for the findings of N-body codes.


The question of the nature of the equilibrium properties of these
core-halo structures is thus relevant both in astrophysics and
cosmology and thus one would like to develop a statistical mechanics
approach to describe these systems. However, one must consider that,
from the point of view of statistical physics, self-gravitating
systems present fundamental problems, that are also common to other
long-range interacting systems.  Indeed, it is well known since the
pioneering works of Boltzmann and Gibbs, that systems with a pair
potential decaying with an exponent smaller than that of the embedding
space, present several fundamental problems that prevent the use of
equilibrium statistical mechanics: thermodynamic equilibrium is never
reached and the laws of equilibrium thermodynamics do not apply
\citep{Padmanabhan_1990,Dux03,Campa08}. Rather these systems reach,
driven by a mean-field collisionless relaxation dynamics,
quasi-equilibrium configurations, or quasi-stationary state (QSS),
whose lifetime diverge with the number of particles $N$
\citep{Dux03,Campa08,yamaguchi,joyceetal_qss,Joyce_Worrakitpoonpon,
tirawut-joyce}.
The formation of QSS is at present one of the most living subjects in
non-equilibrium statistical physics and a general theoretical
framework is still lacking: it is thus necessary to consider toy
models and/or relatively simple systems that can be studied through
numerical well-controlled experiments.


In order to understand the formation of a core-halo structure, a
paradigmatic example is represented by the collapse of a spherical,
isolated and uniform cloud of $N$ randomly placed particles with mass
density $\rho_0$ interacting only by Newtonian gravity. This system
has been considered since the early numerical studies
\citep{henon_1964,vanalbada_1982} when it was realized that it relaxes
violently, in a typical time scale $\tau_D =\sqrt{3\pi/(32G \rho_0)}$,
to produce a virialized state.  Such a time scale is much shorter than
the two-body collisional time scale $\tau_2 \approx N/\log(N) \tau_D$
\citep{binney,Saslaw00} and for this reason in the time range $\tau_D
< t < \tau_2$ the system relaxes into a QSS in virial equilibrium.
Then, because of two-body collisions particles can gain some { 
  kinetic} energy and evaporate from the system: on a time scale of
the order of $\tau_2$ the system changes shape because of particles
evaporation.  Simple considerations based on the microcanonical
entropy (see e.g. \cite{Padmanabhan_1990}) imply that at
asymptotically long times, and for a purely Newtonian potential, the
particles will tend to a configuration in which there is a single pair
of particles with arbitrarily small separation, and the rest of the
mass is in an ever hotter gas of free particles so to conserve the
total energy (see e.g. \cite{aarseth_1974,paper1}).


The underlying physical process in the formation of core-halo
structures in the cosmological context is thought to be similar to the
collective relaxation of such a finite and isolated self-gravitating
particle system.  
\cite{lyndenbell}, who named the collective
relaxation process as ``violent relaxation'' made a theoretical
attempt to explain the gravitational collapse by approximating the
temporal evolution as governed by the collisionless Vlasov equation
and thus neglecting binary collisions.  By introducing a
coarse-graining in phase space the equilibrium state is postulated to
be the one that maximizes the entropy computed by counting all the
possible micro-states compatible with the Vlasov-Poisson conservations
laws.  In this context, differently to ordinary thermodynamic
equilibrium states, the statistical properties of the QSS depend on
initial conditions.  The predictions of the Lynden-Bell approach were
however shown to be at odds with the results of numerical experiments
\citep{arad}.  The failure of the theory was attributed to the fact
that the violent relaxation occurs on very fast dynamical time scale
and the system does not have time to explore all of the phase space to
find the most probable configuration \citep{arad+lyndenbell_2005}.
%


It was recently found by \cite{levin_etal_2008} that the Lynden-Bell
approach, {  considering the system confined in a finite box}, is
able to quantitatively predict the one particle phase space
distribution when the out of equilibrium initial state is close to the
virial requirement, i.e.  $-1.2 \ltapprox b_0 \ltapprox -0.8$, { 
  where 
\be 
b_0 = 
\frac{2K_0}{W_0} 
\ee 
is the initial (i.e., at time $t=0$) virial ratio,
  while $K_0$ and $W_0$ are respectively the initial kinetic and
  potential energy.  }
{  The Lynden-Bell prediction in a confining box is named
``cut-off Lynden-Bell'' and the cut-off is physically justified by the
realization that the relaxation must be restricted to a finite region
of space \citep{phc}}.
Outside this range {  of $b_0$ values } the { 
  cut-off} Lynden-Bell distribution is not able to describe the
statistical properties of the resulting QSS
\citep{levin_etal_2008}. {  When the cut-off is taken to infinity
  the Lynden-Bell distribution is made of a fully degenerate Fermi
  core and particles at infinity, without the halo.}

The {  cut-off} Lynden-Bell {  distribution} was found to be
successful to explain properties of QSS formed in one-dimensional
gravitating systems, for initial conditions near the virial
equilibrium \citep{yamaguchi,Joyce_Worrakitpoonpon,tirawut-joyce}.
{  Recently \citet{teles2011} introduced a novel statistical
  mechanical approach that can avoid some of the fundamental
  assumptions of the Lynden-Bell theory, namely ergodicity and
  phase-space mixing which are generally not satisfied for systems
  with long range forces.}

{  An interesting attempt to construct a statistical mechanics
  modeling of the violent collapse was developed in series of papers
  by
  \cite{stiavelli1987,bertin2003,trenti2005,trentibertin2005,trentibertin2006}.
  This provides physically motivated distribution functions derived
  from the Boltzmann entropy conserving mass, energy, plus a third
  quantity {\it Q}. The problem is, in general, to determine to what
  extent the three quantities are indeed conserved during the
  collapse, i.e.  whether the virialized structure formed after the
  collapse have the same number of particles, energy and {\it Q} of
  the initial mass distribution.}


{  Given that the theoretical problem is very difficult, one needs
  to use gravitational N-body simulations as a means to perform simple
  and controlled numerical experiments.  Previous studies of the
  relaxation of an isolated system starting with cold enough initial
  conditions, i.e. $b_0 \ltapprox 0$,
  \citep{henon_1964,vanalbada_1982,aarseth_etal_1988,boily+athanassoula_2006,
    david+theuns_1989,theuns+david_1990,paper1} have shown that the }
system undergoes to a large contraction, reaching a minimal size,
approximately at $\tau_D$, that scales as $r_{min} \sim N^{-1/3}$ with
the number of points $N$ (at fixed volume $V$ and mass density
$\rho_0=m N/V$).  This behavior can be explained by considering the
growth of density perturbations in the collapsing phase
\citep{aarseth_etal_1988,paper1}.  By neglecting boundary effects, one
may treat the problem by using the linear approximation of the
self-gravitational fluid equations in a contracting universe. The
minimal radius results to be of the order of the unique length scale
characterizing the system, i.e., the initial average distance between
nearest neighbors $\ell \approx r_{min} \propto N^{-1/3}$.

It was then shown \citep{paper1} that a fraction of the particles are
ejected from the system because during the collapse phase they gain
enough kinetic energy. The energy ejected grows approximately as
$N^{1/3}$ while the fraction of the mass ejected slowly changes with
$N$. The mechanism of ejection rises from the interplay of the growth
of perturbations with the finite size of the system.  In particular,
particles lying initially in the outer shells of the system develop a
net lag of their trajectories compared with their uniform collapse
ones. This lag propagates into the volume during the collapse phase
and particles in the outer shells gain positive energy by scattering
through a time dependent potential of an already re-expanding central
core. The resulting density profile of the virialized state is
characterized by a power-law profile of the type $n(r) \sim r^{-4}$
for $r>r_c$. {  Interestingly, this same profile was found
  considering several different systems
  \cite{bertin1984,stiavelli1987}. Note that ejection of mass and
  energy implies that the mass and energy of the virialized structure
  are smaller than the total ones, i.e. there is no mass and energy
  conservation in the collapse.}



In this paper we aim of understanding the origin of the $n(r) \sim
r^{-4}$ density profile, {  investigating the properties of the
  initial conditions necessary to obtain such a behavior}.  In
Sect.\ref{violent_mild_relaxation} we briefly review recent studies of
the warm and cold collapse. The first is defined for the case in which
the initial virial ratio is close to $b_0 \approx -1$ while for the
second close to $b_0 \approx 0$.  {  We motivate the physical
  reasons for such a distinction and} we present in
Sect.\ref{Nbody_simulations} the results of some N-body simulations
where we used the same number of particles but we have varied $b_0$ in
the range $[-1,0]$, with uniform space and velocity distributions
(i.e., water-bag initial conditions).  We show that there is a clear
differences between the structures formed when $b_0 < b_0^c \approx
-1/2$ and $b_0>b_0^c$. We refer to these two relaxation processes,
respectively, as mild and violent: in the latter case the system
contracts by a large factor, while in the former it approximately
maintains its original size.  In Sect.\ref{mild} we discuss in detail
the case of mild relaxation showing that the predictions of the
Lynden-Bell theory {  with a cut-off} agree well with simulations.
Then in Sect.\ref{violent_relax} we show that the main feature of the
$b_0>b_0^c$ case is the $n(r) \sim r^{-4}$ density profile, i.e. the
formation of a dense core and a dilute halo described by such a
power-law profile. In order to explain the origin of this profile we
introduce a simple and well-motivated physical model in
Sect.\ref{model}. Then we discuss (Sect.\ref{critical_value}) the
origin of the critical value $b_0^c \approx -1/2$.  Finally we draw
our main conclusions in Sect.\ref{conclusion} briefly discussing the
relation with the halo structures observed in cosmological N-body
simulations.


\section{Violent and mild relaxation}
\label{violent_mild_relaxation}

As already mentioned, the properties of the QSS resulting from the
collapse of an isolated self-gravitating, spherical, uniform cloud of
particles depend on the initial conditions. In the literature there
have been mostly studied two different cases, i.e.  with initial
virial ratio $b_0\approx-1$ and $b_0=0$, that we are now going to
review in this section.

\subsection{Lynden-Bell theory in a confining box} 
\label{lyndenbell_theory}

Gravitational systems do not reach a time independent equilibrium in
the thermodynamics sense.  Thus the fine-grained distribution function
of positions $\vec{r}$ and velocities $\vec{v}$,
$f(t,\vec{v},\vec{r})$, never reaches a stationary state.
\cite{lyndenbell} {  developed an approach based} on the idea that a
coarse-grained distribution function
$\overline{f}(t,\vec{v},\vec{r})$, averaged on microscopic length
scales, relaxes to a meta-equilibrium distribution
$\overline{f}(\vec{v},\vec{r})$.  The statistical properties of such a
state, differently from the ordinary equilibrium state characterized
by a Maxwell-Boltzmann distribution, explicitly depend on the initial
distribution $f_0(\vec{v},\vec{r})=f(t=0,\vec{v},\vec{r})$.
Lynden-Bell argued that the collisionless relaxation should lead to
the density distribution of levels which is most likely, i.e. the one
that maximizes the coarse-grained entropy, consistent with the
conservation of energy, momentum and angular momentum.

If the initial distribution is a water-bag, i.e. positions are
constrained in $\vec{r} \in [0,\vec{R_0}]$ and velocities in $\vec{v}
\in [0,V_0]$, i.e., 
\be
\label{waterbag}
f_0(\vec{v},\vec{r}) = \eta_1 \Theta(R_0-r) \Theta(V_0-v)
\ee
where $\Theta(x)$ is the Heaviside step
function and $\eta_1=\eta_1(R_0,V_0)$ is a constant, 
the maximization procedure gives a
Fermi-Dirac distribution \citep{levin_etal_2008}
\be
\label{lb1}
\overline{f}(t,\vec{v},\vec{r}) = \eta_1 \rho(\vec{v},\vec{r}) 
= \frac{\eta_1}{\exp 
\left[ \beta ( \epsilon(\vec{v},\vec{r}) -\mu) \right] + 1}
\ee
where $\epsilon(\vec{v},\vec{r})$ 
is the mean energy of particles, $\beta$ and $\mu$ 
are two Lagrange multipliers required by the conservations of energy
and the number of particles, 
\bea
\label{lb2}
&&
\int d^3 r d^3 \overline{f}(t,\vec{v},\vec{r}) 
\epsilon(\vec{v},\vec{r}) = \epsilon_0
\\ \nonumber &&
\int d^3 r d^3 \overline{f}(t,\vec{v},\vec{r}) =1 
\eea 
where $\epsilon_0$ is the energy per particle of
the initial distribution.
{  In this context, 
the incompressibility of the Vlasov dynamics plays the same 
role of the Pauli exclusion principle (see e.g., \cite{phc}).}
Then, the  density profile is simply 
\be
\label{lb4}
n(r) = N \int \overline{f}(\vec{v},\vec{r}) d^3v . 
\ee


In practice, however, what is found is that self-gravitating systems
usually relax to structures characterized by dense cores surrounded by
dilute halos, the distribution functions of which are quite different
from Lynden-Bell $\overline{f}(\vec{v},\vec{r})$.  The failure of the
theory was attributed to the fact that the violent relaxation occurs
on very fast dynamical time scale and the system does not have time to
explore all of the phase space to find the most probable
configurations \citep{arad+lyndenbell_2005}. Numerical simulations,
starting from out of equilibrium configuration characterized by an
initial virial ratio of $b_0 \approx -1/2 $ also showed that
the Lynden-Bell theory, as well as other theoretical attempts, are ad
odds with numerical results \citep{arad}.

However recently \cite{levin_etal_2008} showed that when the initial
distribution satisfies the virial condition $b_0 \approx -1$
the system quickly relaxes to a QSS described quantitatively by the
Lynden-Bell distribution with a cut-off. The cut-off originates from
the requirement that particles must be confined in a finite volume of
space. The reason for this comes from the fact that the possible
configurations include those in which the mass is distributed
throughout space and such a configuration dominates the entropy. The
Lynden-Bell prediction in a confining box is referred as ``cut-off
Lynden-Bell''.
It was then shown that for short enough time scales the precise value
of the cut-off is unimportant \citep{levin_etal_2008}. The
metastable Lynden-Bell distribution persists until a fraction of the
particles evaporates because of two-body collisions.  

{  A similar} agreement {  between the cut-off Lynden-Bell
  distribution and numerical simulations} was also found for initial
conditions close enough to the virial condition, i.e.  $-1.2 \le b_0
\le -0.8$ while outside this range the situation drastically changes
and the Lynden-Bell distribution is not able to describe the
statistical properties of the resulting QSS. Particularly, this occurs
when a fraction of the particles can gain enough kinetic energy to be
ejected from the system in a short time scale, while another part,
which remains bound, form a dense central core and a dilute halo.  This
latter problem is addressed in the following section.

It is interesting to note that, when the cut-off of the truncated
Lynden-Bell distribution is extended to infinity, then the
distribution function splits into two domains, a compact core with
zero temperature plus an evaporated fraction of zero energy particles
at infinity. The distribution function of the core is given by that of
a fully degenerate Fermi gas \citep{levin_etal_2008}. A detailed
comparison of the Lynden-Bell theory, including density profiles,
velocity and energy distributions, with numerical simulations in one
and three spatial dimensions is presented in \cite{tirawut}.


\subsection{Mass and energy ejection}
\label{ejection}

In this section we briefly summarize the main findings by
\cite{aarseth_etal_1988,boily+athanassoula_2006,boily_etal_2002,paper1}
concerning the collapse of a cold, uniform and spherical cloud of
self-gravitating particles.  In the idealized limit of an exactly
uniform spherical distribution different shells do not overlap during
the collapse.  The radial position $r(t)$ of a test particle initially
at $r_0$ is simply given by the homologous rescaling \be
\label{scm1}
r(t) = R(t) r_0 
\ee 
where the scale factor $R(t)$ may be written in the standard parametric form
\bea
&& 
R(\xi) = \frac{1}{2} (1 + \cos(\xi))
\\ \nonumber 
&&
t(\xi) 
= \frac{\tau_{D}}{\pi} \left( \xi + \sin(\xi) \right)
\;, 
\label{scm2}
\eea 
and
\be 
\label{tauscm}
\tau_{D} \equiv \sqrt{\frac{3\pi}{32 G \rho_0}} \;.  
\ee
Eqs.\ref{scm1}-\ref{scm2} describe the unperturbed spherical collapse
model (SCM) trajectories.  At the time $\tau_{D}$ the system collapses
into a singularity.  In a physical situation the collapse is
regularized by perturbations which are present in the initial
conditions at any finite $N$.  At first approximation, one may neglect
the effect of the boundaries on the evolution of the density
perturbations, i.e.  one can consider the limit of an infinite (i.e.,
$R_0 \rightarrow \infty$) contracting system \citep{paper1}. One can
then consider the fluid limit and solve the appropriate equations
perturbatively as it is usually done in cosmology for an expanding
(rather than contracting as in this case) universe \citep{peebles}. A
more detailed approach was developed by \cite{aarseth_etal_1988}
{  taking} explicitly into account the system finite size.

When particles are initially randomly distributed (i.e., with Poisson
fluctuations) one finds that during the collapse the structure reaches
a minimal radius which scales as
\citep{aarseth_etal_1988,boily+athanassoula_2006,boily_etal_2002,paper1}
\be
\label{rmin}
r_{min} \propto N^{-1/3} \;.
\ee 
This scaling with $N$ is obtained by simply taking the criterion
that the SCM breaks down when fluctuations at a
scale of order of the size of the system go non-linear.  Eq.\ref{rmin}
has a very simple interpretation.  Neglecting the finite size of the
system, and given that gravity has no intrinsic length scale, on
purely dimensional grounds we have that $r_{min}$ should be
proportional to the only length scale in the problem, the mean
inter-particle distance $\ell \propto N^{-1/3}$.  Eq.\ref{rmin} has
been observed in N-body simulations by
\cite{aarseth_etal_1988,boily+athanassoula_2006,boily_etal_2002,paper1}.

It was {  then} noticed by \cite{paper1} {  that,} while all
particles start with a negative energy, {  after the collapse} a
finite fraction ends up with positive energy which may escape from the
system.  This transfer of energy occurs in a very short time around
$\tau_D$ and depends on $N$; scaling {  behaviors} with the number
of particles are manifested by the {  amount of ejected energy and
  particles}.  Eq.\ref{rmin}, together with some simple approximations
which have been tested to be valid in the simulations, is the key
element to understand the observed scaling behaviors.

The radial density profile of the virialized structure formed by bound
particles after the collapse was found to have the functional form
\citep{paper1}
\be
\label{dp_b0}
n(r) = \frac{n_c} { 1+ \left( \frac{r}{r_c} \right)^\zeta} \;, 
\ee
where $r_c$ and $n_c$ are parameters depending on $N$ and $\zeta =4$
in agreement with
\cite{henon_1964,vanalbada_1982,stiavelli1987,bertin2003,roy+perez_2004}.
Simple scaling arguments show that $r_c \propto N^{-1/3}$ and $n_c
\propto N^{2}$.  In addition it was also noticed that $r_c \approx
r_{min}$.

Concerning the mechanism of mass ejection it was found that there is a
very clear systematic correlation between particles initial radial
position and ejection, a fact that has lead to understand that the
physical mechanism of ejection indeed arises from the coupling between
the evolution of perturbations and the finite size of the system
\citep{paper1}.  Given the importance of such a mechanism for the rest
of the paper, let us describe it in some details.

The key to understand the ejection mechanism is to realize that
particles initially lying in the outer boundary lag behind the others
during the collapse.  This lag can be understood as follows. Local
density fluctuations modify the SCM trajectories (i.e.,
Eqs.\ref{scm1}-\ref{tauscm}) so that the contraction is no more
perfectly homologous.  In this situation there is an asymmetry between
the shell at the outer boundary compared to {  the ones} in the
bulk: as particles move around there is no compensating inward flux at
the boundary for the mass which moves out under the effect of
perturbations.  For this reason the density of the outer shell
decreases, and also the average density in the sphere at the
corresponding radius, slowing its fall towards the origin. As time
goes on this asymmetry propagates into the volume and for this reason
particles in the outer shell particles arrive at the center of mass on
average much later than those in the bulk.

The mechanism of the gain of energy leading to ejection
is simply that the outer particles, arriving later on 
average, move through the time dependent {\it decreasing}
mean field potential produced by the re-expanding inner mass.
It is possible to work out a  simple estimate for the ejected energy 
that agrees  quite well with the observed scaling  \citep{paper1}.

With respect to the predictions of the theoretical model introduced by
\cite{lyndenbell}, it is interesting to note that, because of
ejection, energy and mass are not conserved during the collapse.  As
discussed in Sect.\ref{lyndenbell_theory} this situation violates the
energy/mass constraints on the final state that is assumed in the
Lynden-Bell treatment. For this reason, it is not surprising that this
approach cannot successfully explain the statistical properties of the
resulting virialized structure.


\section{N-body simulations}
\label{Nbody_simulations}

\subsection{Initial conditions}
\label{initial_conditions}

The initial conditions {  of the simulations} are generated as
follows.  We randomly distribute $N$ particles, of mass $m$, in a
sphere of radius $R_0$ with mass density $\rho_0= 3N/(4\pi R_0^3)
\cdot m $\footnote{Our units are such that $\rho_0=1$ gr/cm$^3$ so
  that $\tau_D=2100$ seconds}.  The gravitational potential energy at
time $t$ is
\be
W(t) = - \frac{1}{2} \sum_{i=1}^{N}\sum_{j=1}^{N}  \frac{G m_i m_j}{r_{ij}}\,. 
\ee
where $r_{ij}$ is the distance of the $i^{th}$ from the $j^{th}$ particle.
The total kinetic energy is simply 
\be
\label{kinetic_energy}
K(t) = \frac{1}{2} m \sum_{i=1}^{N} v_i(t)^2 \, 
\ee
where $v_i(t)$ is the velocity of the $i^{th}$ particle.
The virial ratio is 
\be
\label{virial_ratio} 
b(t) = \frac{2 K(t)}{W(t)} \;. 
\ee

We generate a series of spherical clouds of particles, with $N=10^4$
and with different initial virial ratio $b_0=b(t=0)$.  We
take the velocity components to have a uniform probability density
function (PDF) in the range $[-V_0,V_0]$, and the modulus of the
velocity is constrained to be in a sphere of radius $V_0$.  The
velocity PDF is thus 
\be g(v) = \frac{3}{V_0^3} v^2 \;\; \mbox{for}
\;\; v \le V_0 
\ee 
and zero otherwise. Such a PDF clearly satisfies
\be 
\int_0^{\infty} g(v) dv = \int_0^{V_0} g(v) dv = 1 \;. 
 \ee 
The
initial velocity dispersion is \be \langle v^2 \rangle = \int_0^{V_0}
v^2 g(v) dv = \frac{3}{5}V_0^2 \ee where {  we defined}
\be 
\label{v02}
V_0^2 = \frac{ b_0 G N  m}{R_0} \;. 
\ee 
{  To obtain Eq.\ref{v02} } we used that the gravitational potential
energy of a uniform spherical mass distribution is \citep{binney} \be
\label{potential_energy}
W_0= - \frac{3}{5} \frac{ G(mN)^2 } {R_0} \; .
\ee
The initial conditions are thus constrained in a {\it water-bag} distribution.


\subsection{Code and numerical parameters}

To run N-body simulations we have used the parallel version of the
publicly available tree-code GADGET
\citep{springel_2005,gadget_paper}.  There are various parameters of
the code that must be tuned in order to have a good accuracy in the
time integration: {  as a control we have used both energy and
  angular momentum conservation, which are a sensitive monitoring of
  the accuracy of the simulation \citep{aarseth_book}}.  We used a
force softening such that $\epsilon/\ell = 0.007$ where $\ell \approx
0.55 (4\pi R_0^3/3N)^{1/3}$ is the initial inter-particle distance.
Note that the minimal radius $r_{min}$ of the structure in the case of
maximum contraction, i.e. when $b_0=0$, is found to be $r_{min}
\approx \ell$ (see Sect.\ref{ejection}).  As discussed in
\cite{paper1}, where a number of tests with different values of
$\epsilon$ were performed, the dynamics of the collapse phase and the
formation of the QSS remains unchanged as long as $\epsilon<
\ell,r_{min}$.

In addition to the softening length, the accuracy of a 
GADGET simulation is determined by the internal time-step accuracy and by 
the cell-opening accuracy parameter of the force calculation We chose
the time-step criterion 0 of GADGET with $\eta=0.01$. In the force
calculation we employed the new GADGET cell opening criterion with a
high force accuracy of $\alpha_F = 0.001$
\citep{springel_2005,gadget_paper}.

\begin{figure}
\vspace{1cm}
{
\par\centering \resizebox*{9cm}{8cm}{\includegraphics*{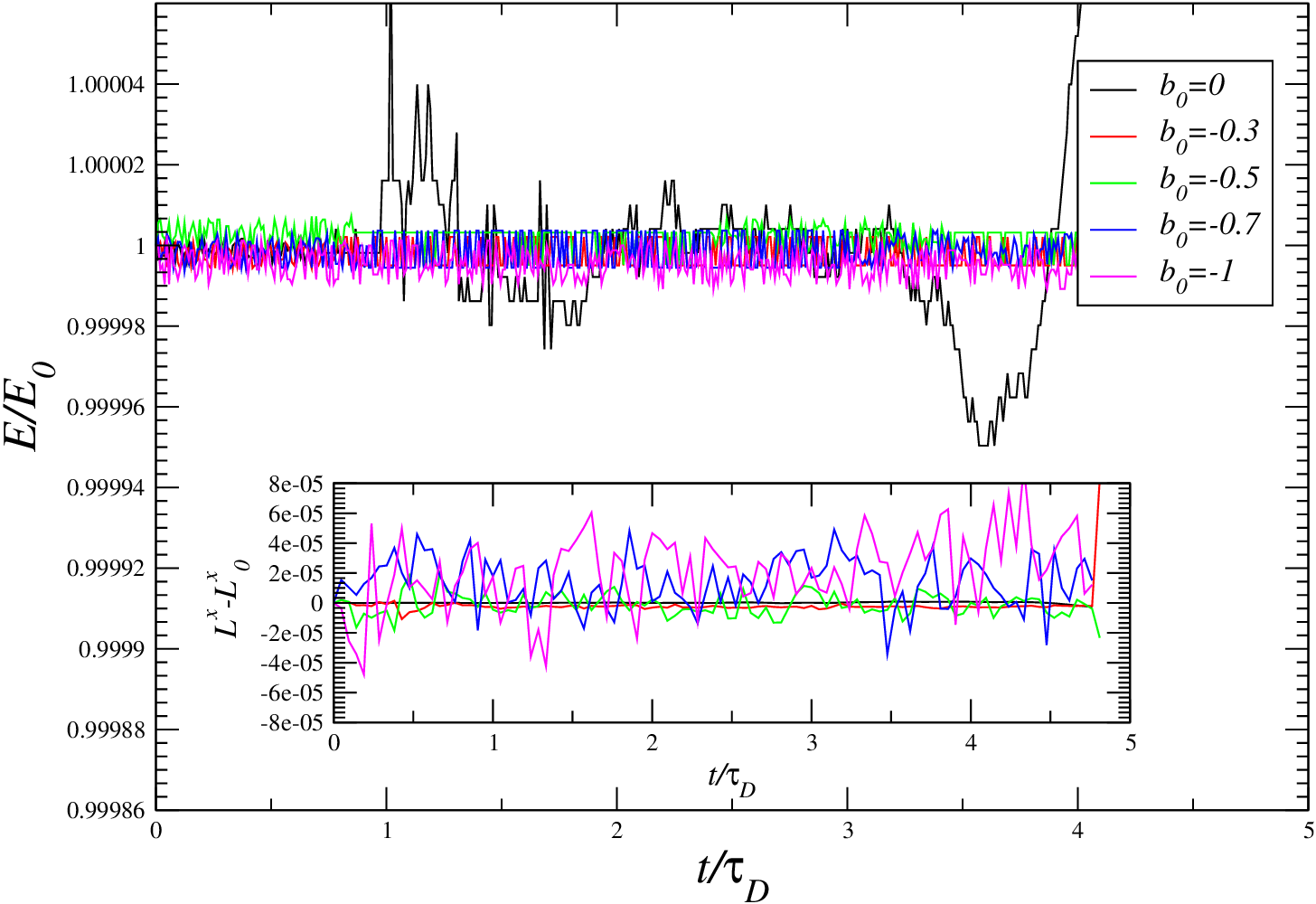}}
\par\centering
}
\caption{Behavior of the total energy {  normalized to its initial
    value} as a function of time for different values of $b_0$. { 
    In the inset panel it is shown the behavior of one of the
    components (i.e., $L^x$) of the total angular momentum as a
    function of time}.}
\label{Energy_cons}
\end{figure}

The behavior of the energy conservation is shown in
Fig.\ref{Energy_cons}: we have that $\Delta E(t) / E_0 \ll 5 \times
10^{-3}$ (where $E_0=W_0+K_0$ is the initial total energy\footnote{
  {  In the computation of the gravitational potential energy we
    have taken into account the shape of the gadget softened potential
    \citep{springel_2005,gadget_paper}.}}) when $b_0=0$, in the range
of time we have considered $0 \le t \le 4 \tau_D$; in the other cases
energy conservation is about {  $\sim 10^{-3} $}. One may note
that the larger is $b_0$ the less accurate is energy conservation as
the system size gets smaller and particles gain higher velocities. The
latter is the reason for the largest deviation in the energy
conservation seen for $b_0=0$ at $t \approx 4 \tau_D$.  {  Moreover,
  the behavior as a function of time of one component (for instance
  along the x-axis) of the total angular momentum shows that it is
  well conserved during the time integration (see inset panel of
  Fig.\ref{Energy_cons}).  }

\subsection{Global behaviors} 

\begin{figure}
\vspace{1cm} { \par\centering
\resizebox*{9cm}{8cm}{\includegraphics*{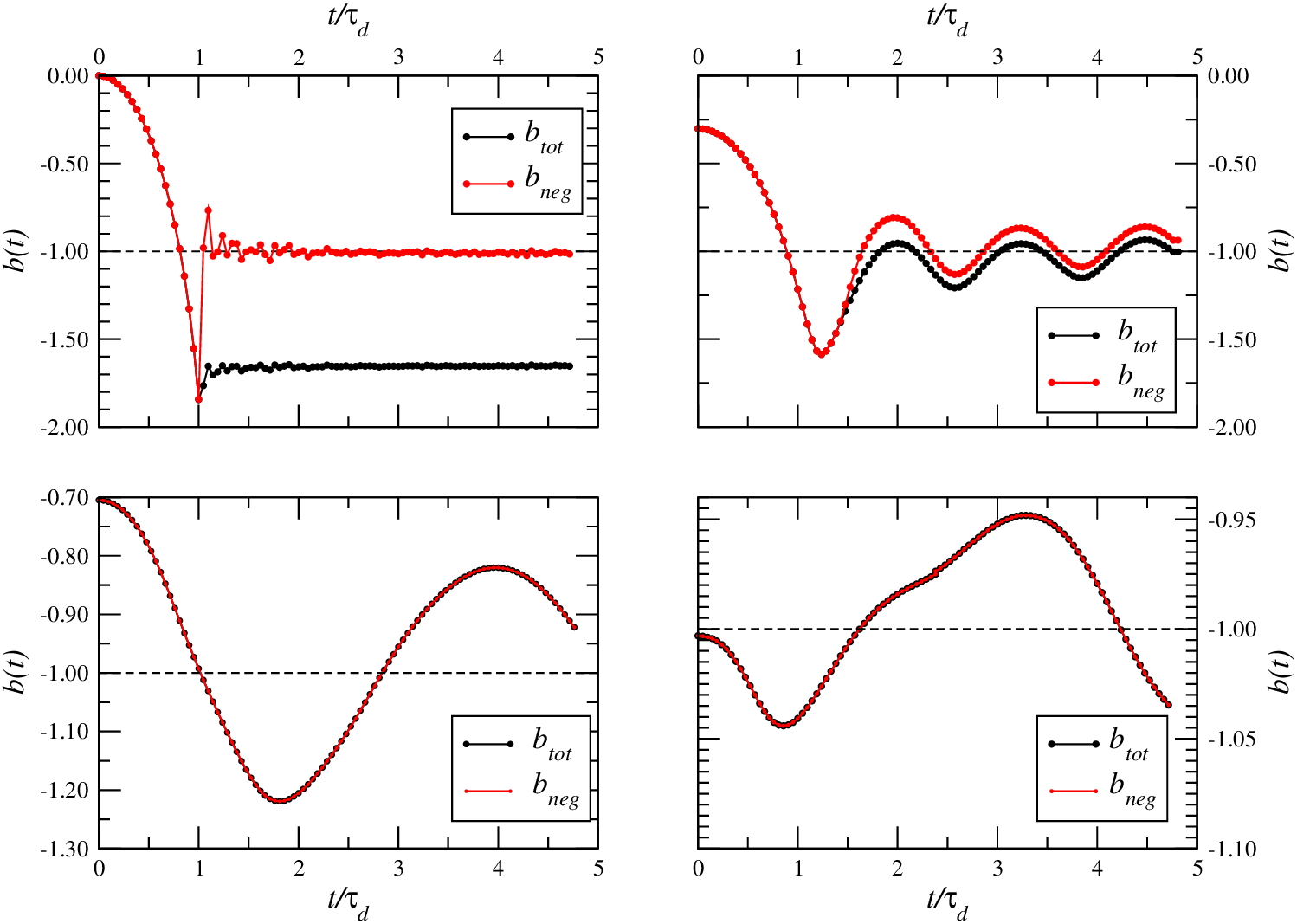} }
  \par\centering }
\caption{Behavior of the virial ratio for all system particles (black
  line) and only for particles with negative total energy (red
  line) as a function of time for different values of the
  initial virial ratio. { \it Upper left panel:} $b_0=0$, 
  { \it upper right panel:} $b_0=-0.3$,
  { \it bottom left panel:} $b_0=-0.7$ and 
  { \it bottom right pane:l} $b_0=-1$.
   \label{Virial_total} }
\end{figure}

The virial ratio as a function of time $b(t)$ shows a different
behavior depending on $b(t=0)=b_0$ (see Fig.\ref{Virial_total}). For
$b_0<-1/2$, $b(t)$ presents a series of damped oscillations around the
asymptotic value $-1$. Instead, for $ b_0 =0 $ it presents a sharp
change of behavior at $\tau_D$.  In addition, one may note that, for
$t > \tau_D$, the virial ratio of the fraction of particles with
negative total energy stabilizes, as expected, around $b_{neg} \approx
-1$, while the virial ratio of all the $N$ system particles reaches
the an asymptotic value that is $b_{tot} < b_{neg}$.

This behavior is easily explained by considering the ejection of a
fraction of the particles from the system --- i.e., for $b_0 > -1/2$ a
certain fraction of the particles gain positive energy during the
collapse.  Their kinetic energy is the origin of the offset between
$b_{tot}$ and $b_{neg}$.  {  Indeed,} the potential energy of the
particles with positive energy becomes negligible (i.e., $|W_{pos}|
\ll |W_{neg}|$) because their distance from the structure rapidly
increases, so that at first approximation we have
\be
b_{tot} = \frac{2 K_{tot}}{W_{tot}} \approx b_{neg} + \frac{2
  K_{pos}}{W_{neg}} < b_{neg}  \;.
\ee 
On the other hand, for $b_0 < -1/2$
all particles remain bounded to the structure and thus 
$b_{neg}(t)=b_{tot}(t)$.

\begin{figure}
\vspace{1cm}
{
\par\centering \resizebox*{9cm}{8cm}{\includegraphics*{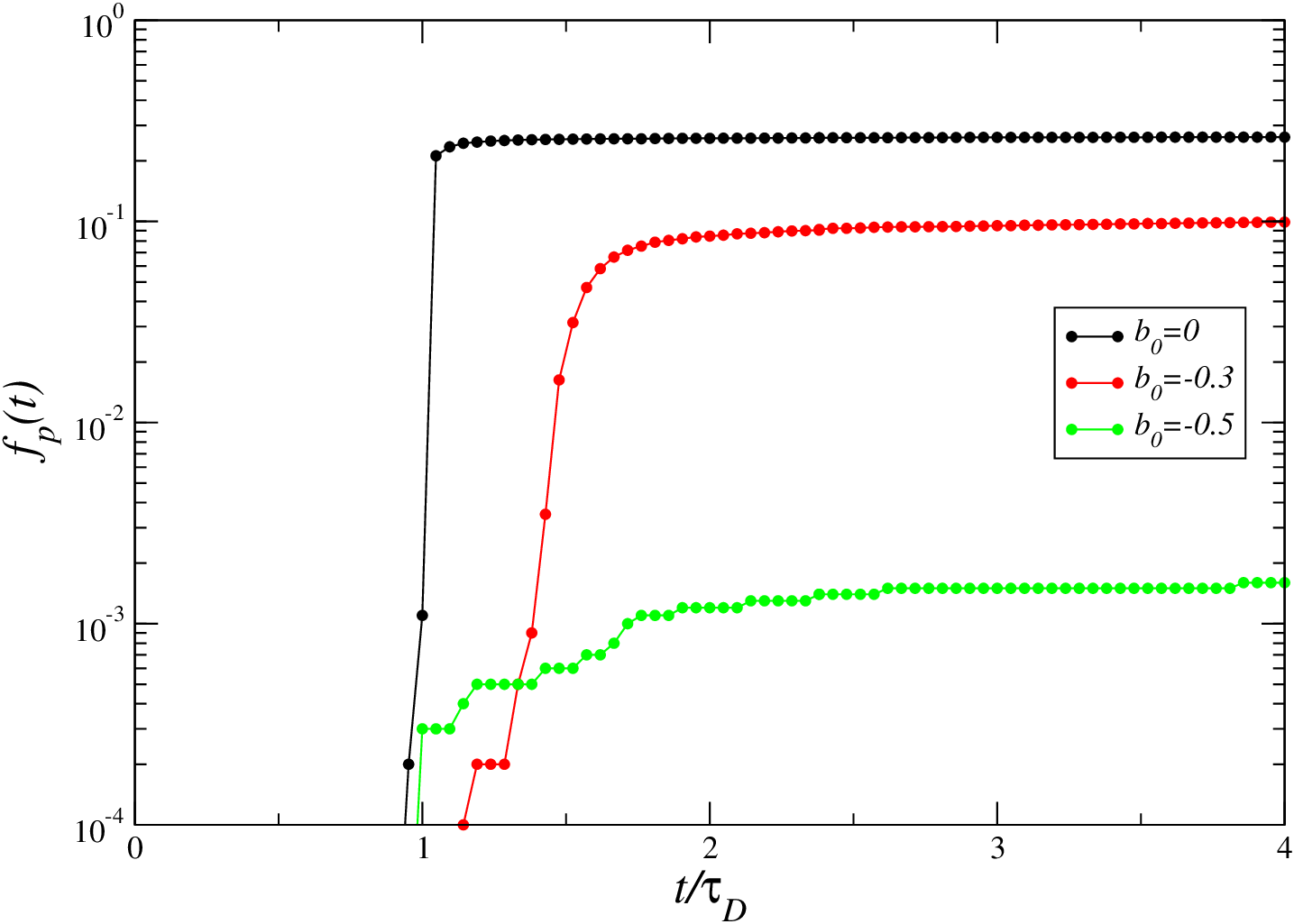} }
\par\centering
}
\caption{Fraction of the particles with positive energy function of
  time for different values of $b_0$: for $b_0 < -1/2$ there is no
  ejection of particles.}
\label{fp_time} 
\end{figure}

{  This picture is conformed by Fig.\ref{fp_time} that} shows the
fraction of particles $f_p(t) $ with positive energy as a function of
time: we find $f_p(t) >0$ for $t> \tau_D$ and $b_0 > -1/2$. {  On
  the other hand,} for $b_0 < -1/2$ there is no ejection and $f_p(t)=0
\;\; \forall t$.

\begin{figure}
\vspace{1cm}
{
\par\centering \resizebox*{9cm}{8cm}{\includegraphics*{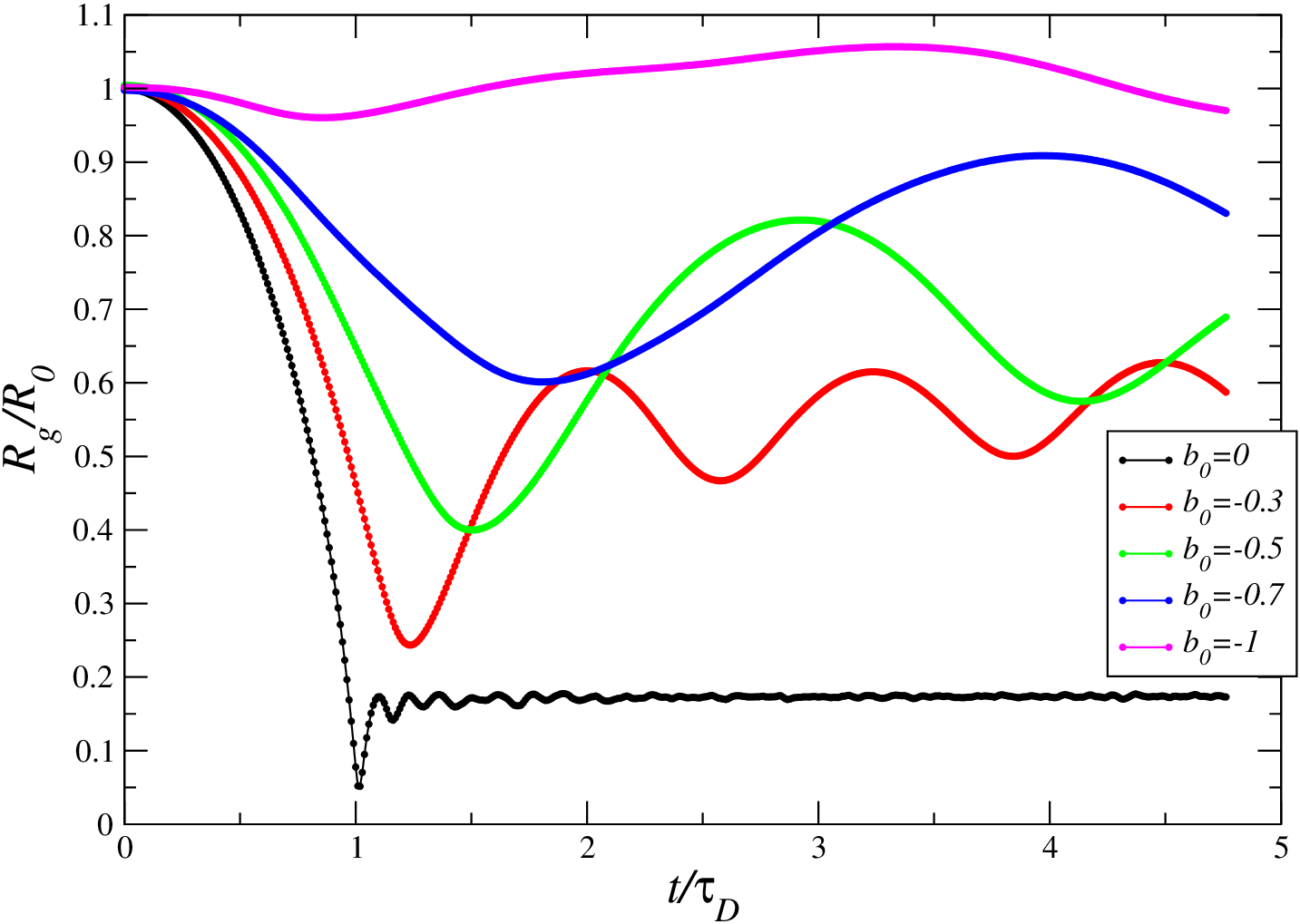}}
\par\centering
}
\caption{Gravitational radius of the structure as function of time for 
different initial virial ratio $b_0$.}
\label{R0t}
\end{figure}

As long as the spherical structure has uniform density
the gravitational radius 
\be R_g(t) = - \frac{3}{5}
\frac{GM^2}{W(t)} \;,
\ee 
coincides with the physical radius.  From the analysis of the
behavior of $R_g(t)$ shown in Fig.\ref{R0t} we may conclude that
minimal size of the structure also depends on $b_0$.  In particular,
the minimal size $r_{min} \ll R_0$ is reached when $b_0 \rightarrow
0$.  while for $b_0 < -1/2$ the size of the structure is almost
unchanged.

In summary we have found that there is a clear difference between the
behaviors of the relevant physical quantities for different initial
virial ratio, particularly when the $b_0$ is smaller or larger than
$b_0^c \approx -1/2$.  In what follows we will study the statistical
properties of the resulting quasi-equilibrium structure: we firstly,
in Sect.\ref{mild}, discuss the problem of ``mild relaxation'',
i.e. $b_0 < b_0^c $, to then pass in Sect.\ref{violent_relax}  to the
problem of ``violent relaxation" for $b_0 > b_0^c $. In
Sect.\ref{critical_value} we will consider the problem of { 
  understanding the origin} the (approximate) value of $b_0^c$.


\section{Mild relaxation and the Lynden-Bell predictions} 
\label{mild} 

\begin{figure}
\vspace{1cm}
{
\par\centering \resizebox*{9cm}{8cm}{\includegraphics*{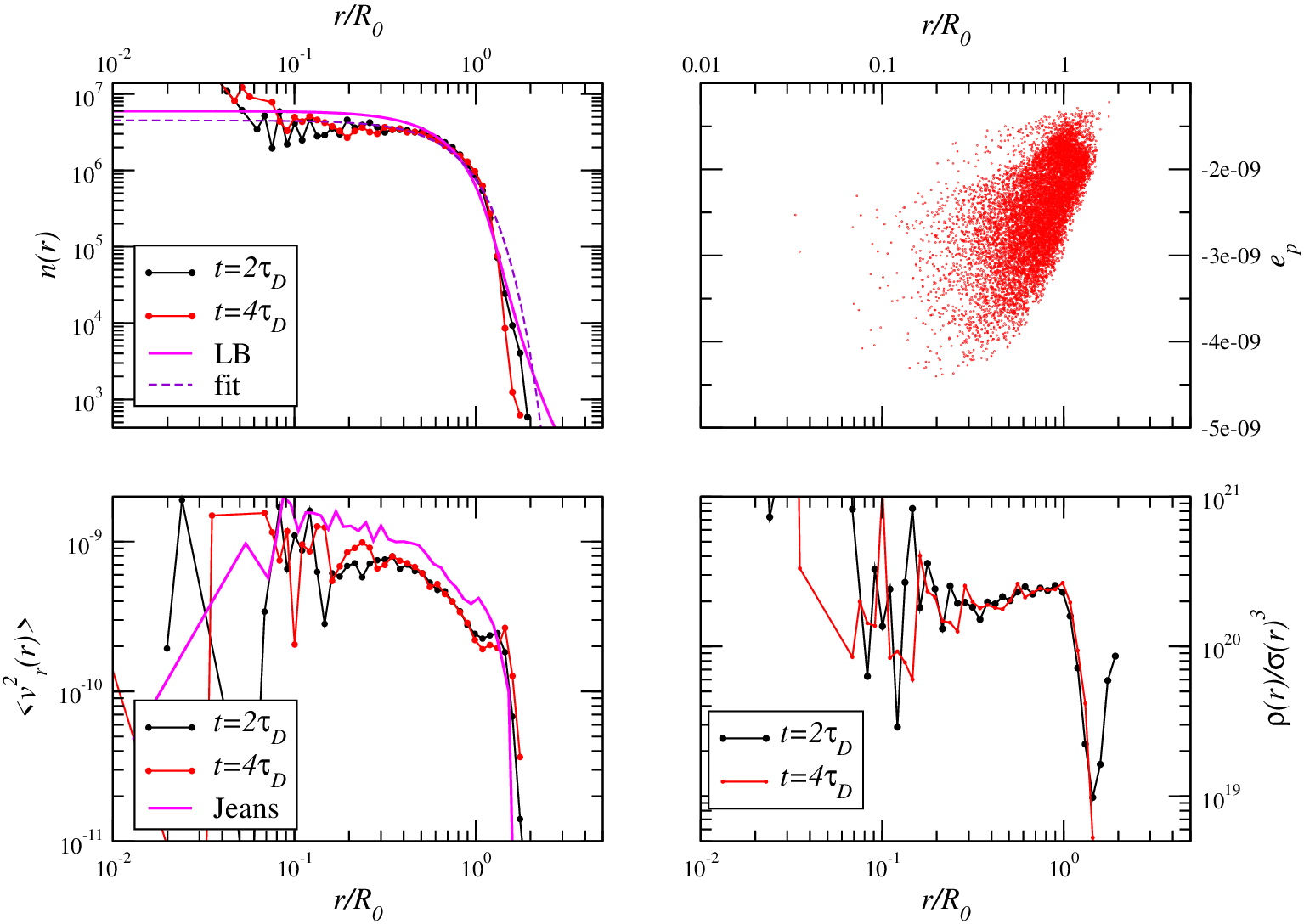}}
\par\centering
}
\caption{Behavior of some statistical quantities for the case
  $b_0=-1$.  Upper left panel: density profile together with the
  prediction of the Lynden-Bell theory (LB) and the best fit with
  Eq.\ref{dp_b1}.  Upper right panel: energy per particle $e_p$ as a
  function of the distance from the center at $t=4\tau_D$.  Bottom
  left panel: mean square value of the radial velocity together with
  the prediction of the Jeans equation (Eq.\ref{Jeans4}).  Bottom
  right panel: phase space density $\rho/\sigma^3(r)$.  }
\label{Fig1_b1}
\end{figure}

Let us firstly discuss the case $b_0=-1$. Hereafter, we identify the
center of the structure as the point in which the potential is
minimum: alternative definitions (i.e. the center of mass) do not
change qualitatively the results discussed below.  The density profile
is shown in Fig.\ref{Fig1_b1} (upper left panel) together with the
{  cut-off} Lynden-Bell {  distribution}\footnote{I thank Yan
  Levin and Renato Pakter for their data on the {  cut-off}
  Lynden-Bell distribution.}  (see Sect.\ref{lyndenbell_theory}),
which nicely fits the measured behavior. A more detailed comparison of
the results of N-body simulations with the predictions of the
Lynden-Bell theory can be found in \cite{tirawut}, where it is
discussed that also the energy and velocity distributions are in good
agreement with the theoretical behaviors.
The density profile can be best-fitted  by a function of the
type 
\be
\label{dp_b1} 
n(r) = n_c \exp(-(r/r_c)^\eta) \;,
\ee
where  $\eta \approx 2$.   In addition we find that the characteristic
length scale $r_c$ is of the same order $R_0$, implying that the system
has not gone through a drastic change of shape and size. Rather it is 
only slightly changed so that particles have rearranged their positions
and velocities to find a quasi-equilibrium configuration. 
In Fig.\ref{Fig1_b1} (upper right panel) it is also shown the behavior
of the energy $e^i_p$ of the $i^{th}$ particle as a function of its
distance from the center. We may note that $e^i_p <0 \;\; \forall i$,
which corresponds to the fact that all particles are bound: {  note
  no clear correlation between energy and spatial position is
  detected.}

For an isotropic radial density profile, $\rho(r)$, 
one may solve, analytically or numerically,  
the Jeans equation to get the corresponding velocity dispersion,
$\sigma^2(r)$ \citep{Hernquist_1990,Tremaine_et_al_1994}. 
{  
The Jeans equation is 
\begin{equation}
\label{Jeans2}
\frac{1}{\rho(r)} 
{{d(\overline{v_r(r)}^2 \rho(r))}\over{dr}} 
+ \alpha(r) \frac{ \overline{v_r(r)}^2}{r} 
= - {{d\Phi}\over{dr}} \;.
\end{equation}
In the previous equation $\sigma(r) = \overline{v_r(r)}^2$ is the
velocity dispersion in the radial direction,  
\be
\label{alpha} 
\alpha(r) = 2 -  \frac{\overline{v_t(r)}^2} {\overline{v_r(r)}^2 } \;,
\ee
is the the anisotropy parameter 
and  $v_t(r)$ is the velocity 
in the transversal direction.
When $\overline{v_t(r)}^2 = \overline{v_r(r)}^2$ 
the velocity anisotropy terms are zero and 
Eq.\ref{Jeans2} can be rewritten as 
\footnote{  A more detailed study of the stationary solutions of the
  Vlasov equation should consider the solution of Eq.\ref{Jeans2} with
  a non-zero anisotropy term (see e.g., \cite{trentibertin2006}).}
\be
\label{Jeans4}
\overline{v_r(r)}^2 = \frac{1}{ \rho(r) } \int_r^\infty \frac{\rho(y)
  G M(y)}{y^2} dy \;, 
\ee 
with the boundary condition 
\be 
\lim_{r  \rightarrow \infty} \overline{v_r(r)}^2 \rho(r) = 0 \;.  
\ee
It is interesting to note that the Jeans equation (Eq.\ref{Jeans4})
is reasonably well satisfied in the time range we consider
(Fig.\ref{Fig1_b1} --- bottom left panel): this implies that the
stationary state is well described by a stationary solution of the
Vlasov equation, i.e. it is a collisionless stationary state. It
should be noticed that although the velocity anisotropy
(Eq.\ref{alpha}) is different from zero (see below), the perturbation
to the Jeans equation due to such a term does not sensibly affect the
agreement between the measured $\overline{v_r(r)}^2$ and
Eq.\ref{Jeans4}. (We will come back on this point in
Sect.\ref{violent_relax}.) }
Finally we note that the
phase-space density $\rho/\sigma^3(r)$, where $\sigma^2 \equiv
\langle v^2_r(r) \rangle$ is about flat, with a sharp decay for $r
\rightarrow R_0$.

\begin{figure}
\vspace{1cm}
{
\par\centering \resizebox*{9cm}{8cm}{\includegraphics*{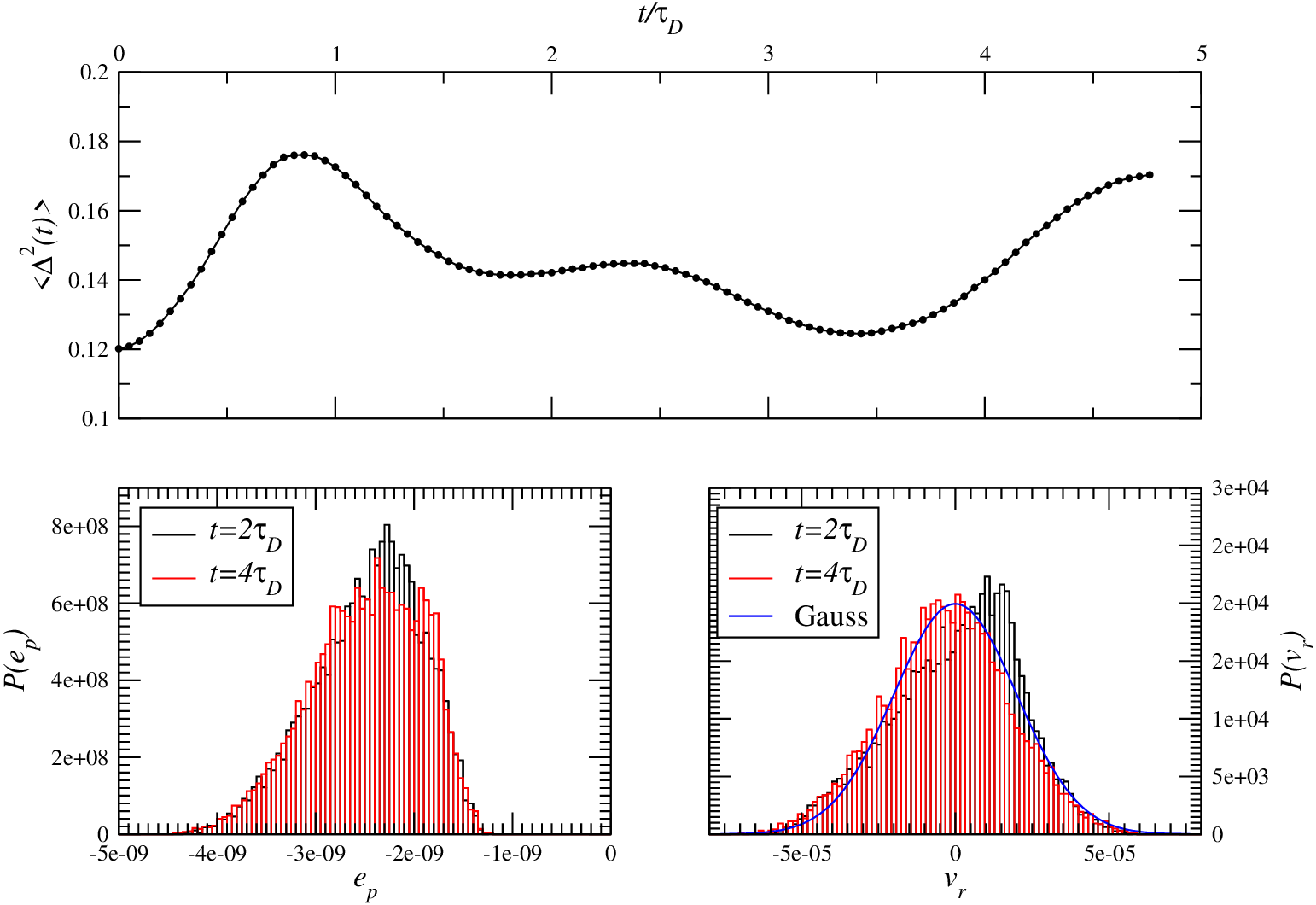}}
\par\centering
}
\caption{Other statistical quantities for the case $b_0=-1$.  {\it
    Upper panel}: particle energy fluctuations (see Eq.\ref{delta2}).
  {\it Bottom left panel}: energy distribution of all particles at
  different times.  {\it Bottom right panel}: radial velocity
  distribution of particles at different times together with a the
  best fir with a Gaussian function.  }
\label{Fig2_b1}
\end{figure}

A statistical measure of the amount of energy that all particles have
exchanged can be defined as follows 
\be
\label{delta2}
\langle \Delta^2 (t) \rangle = \frac{1}{N(N-1)} 
\frac{\sum_{i,j=1 \; i\ne j }^{N} (e_p^i(t) -e_p^j(t))^2 }
{\langle e(t) \rangle^2}
\ee
where $e_p^i(t)$ the average energy per particle is defined as 
\be
\langle e(t) \rangle = \frac{\sum_{i=1}^{N_t} e_p^i(t)}{N} \;. 
\ee 
One may see from Fig.\ref{Fig2_b1} that $\langle \Delta^2 (t) \rangle$
oscillates in phase with the virial ratio (see Fig.\ref{Virial_total})
and that the amount of energy exchanged by all particles is smaller
than $10\%$ during the whole time range considered.

\begin{figure}
\vspace{1cm}
{
\par\centering \resizebox*{9cm}{8cm}{\includegraphics*{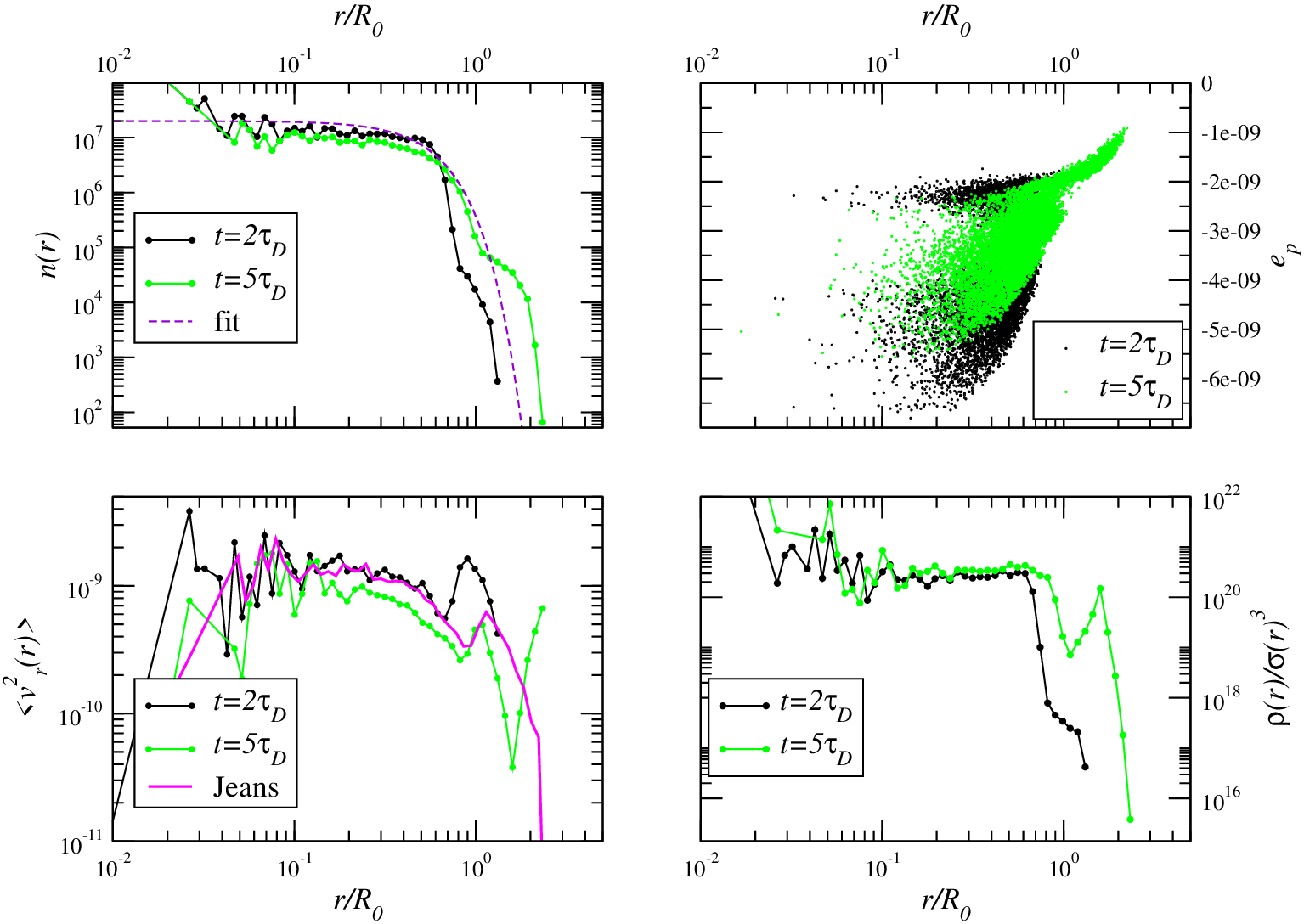}}
\par\centering
}
\caption{As Fig.\ref{Fig1_b1} but for the case $b_0=-0.7$.}
\label{Fig1_b07}
\end{figure}

\begin{figure}
\vspace{1cm}
{
\par\centering \resizebox*{9cm}{8cm}{\includegraphics*{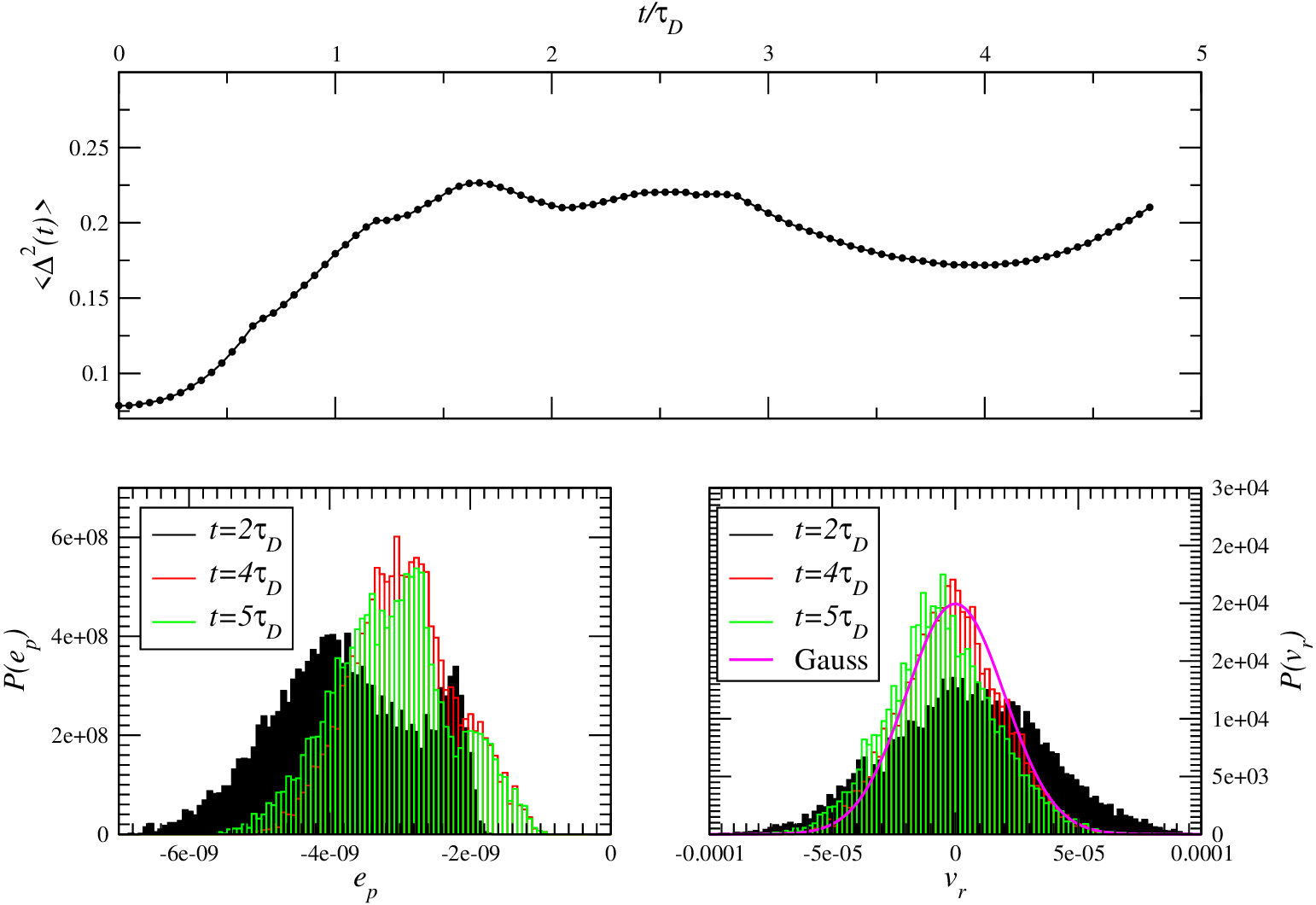}}
\par\centering
}
\caption{As Fig.\ref{Fig2_b1} but for the case $b_0=-0.7$.}
\label{Fig2_b07}
\end{figure}

The case $b_0=-0.7$ does not show substantial differences with respect
to the $b_0=-1$ case (see Figs.\ref{Fig1_b07}-\ref{Fig2_b07}).  The
prediction of the Jeans equation for the velocity dispersion shows
again that the system is well described by the collision-less limit
{  (neglecting the  term $\alpha(r)$ in Eq.\ref{Jeans2})}. 
The density profile is still characterized by a constant behavior at
small scales followed by a sharp decay of the type described by
Eq.\ref{dp_b1}, although $r_c$ is smaller than for the $b_0=-1$ case,
implying a larger contraction during the collapse phase.
Correspondingly, particle energies, for $t> 2 \tau_D$, are 
larger than for the $b_0=-1$ case, but still $e_p^i <0 \; \forall
i$. The exchange of energy among particles (Eq.\ref{delta2}) was more
efficient during the first oscillation of the system, i.e. for $0 < t
< 1.5 \tau_D$, and it is then reduced at later times, in agreement
with the fact that the system is relaxed into a QSS: each particle
move in a time independent potential and the energy of each particle
is conserved modulo two-body collisions.


\section{Violent Relaxation and the formation of the power-law tail 
of the density profile} 
\label{violent_relax} 

We now present the main results of N-body simulations for the case in
which the initial virial ratio is $-1/2 < b_0 \le 0$. In this case
during the collapse the size of the system undergoes to a large
compression and a fraction of the particles gain a certain amount of
kinetic energy so that they will have velocities larger than the
escape one.

\begin{figure}
\vspace{1cm}
{
\par\centering \resizebox*{9cm}{8cm}{\includegraphics*{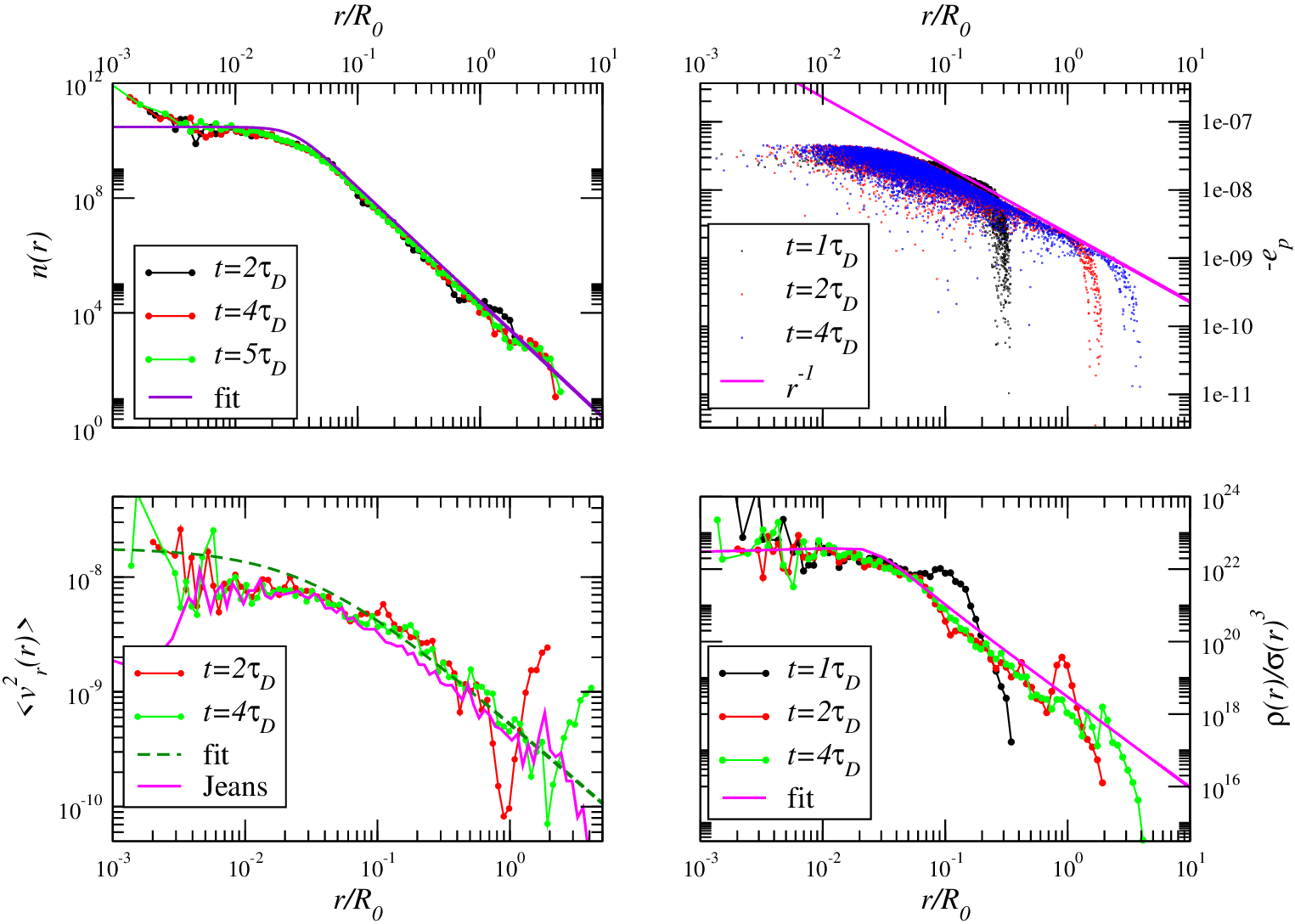}}
\par\centering
}
\caption{As Fig.\ref{Fig1_b1} but for $b_0=0$. The best fit of the
  density profile with Eq.\ref{dp_b0} is also shown, together with the
  fit given by Eq.\ref{fitv2r} of mean square value of the radial
  velocity and the fit given by Eq.\ref{fitpsd} for the phase space
  density.
\label{Fig1_b0}
}
\end{figure}

In Fig.\ref{Fig1_b0} (upper left panel) it is shown the density
profile at $t>\tau_D$: one may note that an almost asymptotic behavior
is reached already at $t \gtapprox \tau_D$.  However, at later times
the profile is almost identical but for the fact that the tail extends
to larger scales.  We find that the density profile is well
approximated by Eq.\ref{dp_b0} where $r_c \approx 0.03 R_0$ and $\zeta
= 4$.  Note that the density profile has two different regimes: at
small scales, i.e. $r<r_c$, the structure corresponds to an
homogeneous sphere with constant density $n(r) \approx n_c$ while at
large scales, i.e. $r>r_c$ it shows a $r^{-4}$ decay.  As already
mentioned we find that $r_c \approx \ell = 0.55 R_0 (4\pi/N)^{1/3}$.
The minimal radius $r_{min}$ reached by the structure during the
collapse can be defined as the radius, measured from the center of
mass, enclosing the $90\%$ of the mass. It is found that $r_{min}
\approx r_c \approx \ell$.

In the upper right panel of Fig.\ref{Fig1_b0} it is shown the diagram
radial distance-total energy only for the bound particles. One
may note that for $r>r_c$ the points follows an approximate $e_p \sim
r^{-1}$ behavior.  This can be easily explained by considering that
particles at distances $r>r_c$ move in a constant gravitational
potential generated by particles with $r<r_c$. In this situation
particle velocities should display a Keplerian behavior $v_r \sim
r^{-1/2}$ so that that $e_p \sim v_r^2 \sim r^{-1}$.  This
behavior is confirmed by considering (bottom right panel of
Fig.\ref{Fig1_b0})  the behavior of the average radial component of
the velocity as a function of the radial distance. (The average has
been performed in radial shells). Indeed, we
find that
\be
\label{fitv2r} 
\langle v_r^2 (r) \rangle \equiv \sigma^2(r) =
\frac{\sigma_c^2}{1+ \left( \frac{r}{r_c} \right)} \;, 
\ee 
where $\sigma_c^2$ is a constant and $r_c$ has been determined from
the density fit (Eq.\ref{dp_b0}).  The behavior of Eq.\ref{fitv2r} is
similar to the one of the density in Eq.\ref{dp_b0}: it is constant at
small scales and it decays for $r> r_c$.  As time passes, a few
particles reach a larger and larger distances from the center of the
structure, leaving however unchanged the functional behavior of
Eq.\ref{fitv2r}.

Given the behaviors of Eq.\ref{dp_b0} and Eq.\ref{fitv2r} we may fit
the phase-space density with (see the bottom left panel of
Fig.\ref{Fig1_b0}) 
\be
\label{fitpsd}
\frac{\rho(r)}{\sigma^3(r)} = \frac{n_c} { 1+ \left( \frac{r}{r_c}
  \right)^4} \times \sqrt{\frac{1+ \left( \frac{r}{r_c} \right)}
  {\sigma_c^2} }^3 \;.  \ee 
Thus we find that the phase-space density is
{  $\rho/\sigma^3 \propto r^{-5/2}$} for $r>r_c$ while it is almost flat
at smaller scales.

\begin{figure}
\vspace{1cm}
{
\par\centering \resizebox*{9cm}{8cm}{\includegraphics*{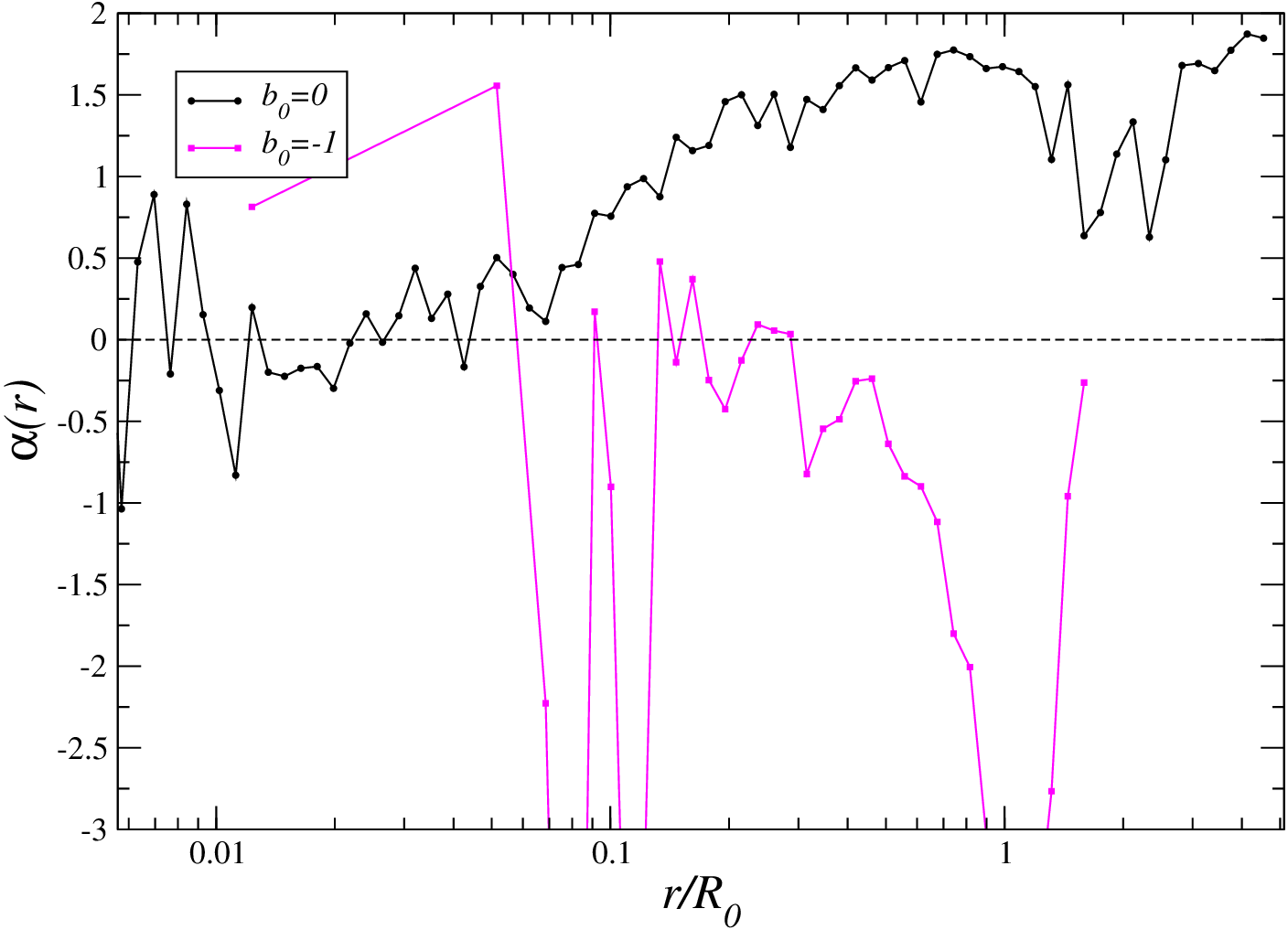}}
\par\centering
}
\caption{{  Velocity anisotropy for the case $b_0=0$ and
    $b_0=-1$. When $\alpha(r)>0$ the velocity dispersion for $b_0=0$ is
    by dominated by its radial component while for $b_0=-1$ by its
    transversal one.}  }
\label{Fig_VelAn}
\end{figure}

{  It is interesting to note, as firstly shown in the pioneering
  paper by \citet{vanalbada_1982} and then studied in detail by
  \citet{trenti2005}, that the QSS formed after the collapse is
  dominated, in the outer regions where the density scales as $n(r)
  \sim r^{-4}$, by radial orbits. This is shown by the behavior of
  $\alpha(r)$ (Eq.\ref{alpha}) as a function of the radial distance
  (see Fig.\ref{Fig_VelAn}).  On the other hand for the QSS obtained
  starting from $b_0=-1$ the velocity dispersion is by dominated by
  its transversal component.}

\begin{figure}
\vspace{1cm}
{
\par\centering \resizebox*{9cm}{8cm}{\includegraphics*{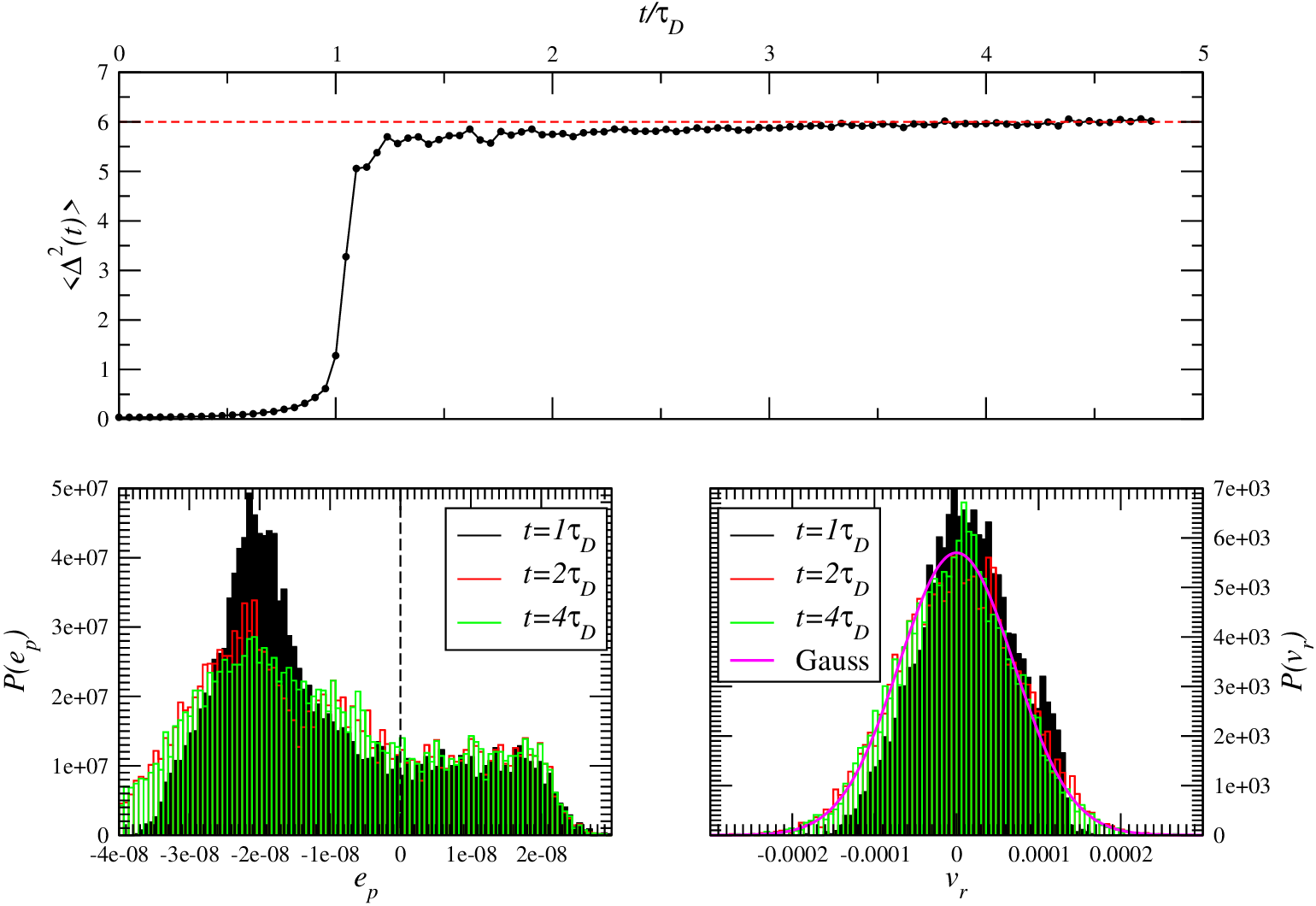}}
\par\centering
}
\caption{As Fig.\ref{Fig2_b1} but for $b_0=0$. 
}
\label{Fig2_b0}
\end{figure}

The behavior of $\Delta^2(t)$ (Eq.\ref{delta2}) (see the upper panel
of Fig.\ref{Fig2_b0}) shows that in this case particles exchange a
substantial amount of energy in the time range $0.7 \tau_D < t < 1.2
\tau_D$, while before and after the central collapse phase the energy
per particle is very well conserved.  Differently from the $b_0<-1/2$
case, during the collapse a fraction of the particles change their
total energy by a relevant factor so that some of the particles may
gain enough kinetic energy to escape from the system. The $b_0> -1/2$
case corresponds to an almost instantaneous collapse followed by a
rapid relaxation toward a QSS.  The particle energy distribution
(Fig.\ref{Fig2_b0} bottom left panel) shows that some of the particles
have indeed positive energy.  Finally the radial velocity distribution
(Fig.\ref{Fig2_b0} bottom right panel) is reasonably well fitted by a
Gaussian function.

\begin{figure}
\vspace{1cm}
{
\par\centering \resizebox*{9cm}{8cm}{\includegraphics*{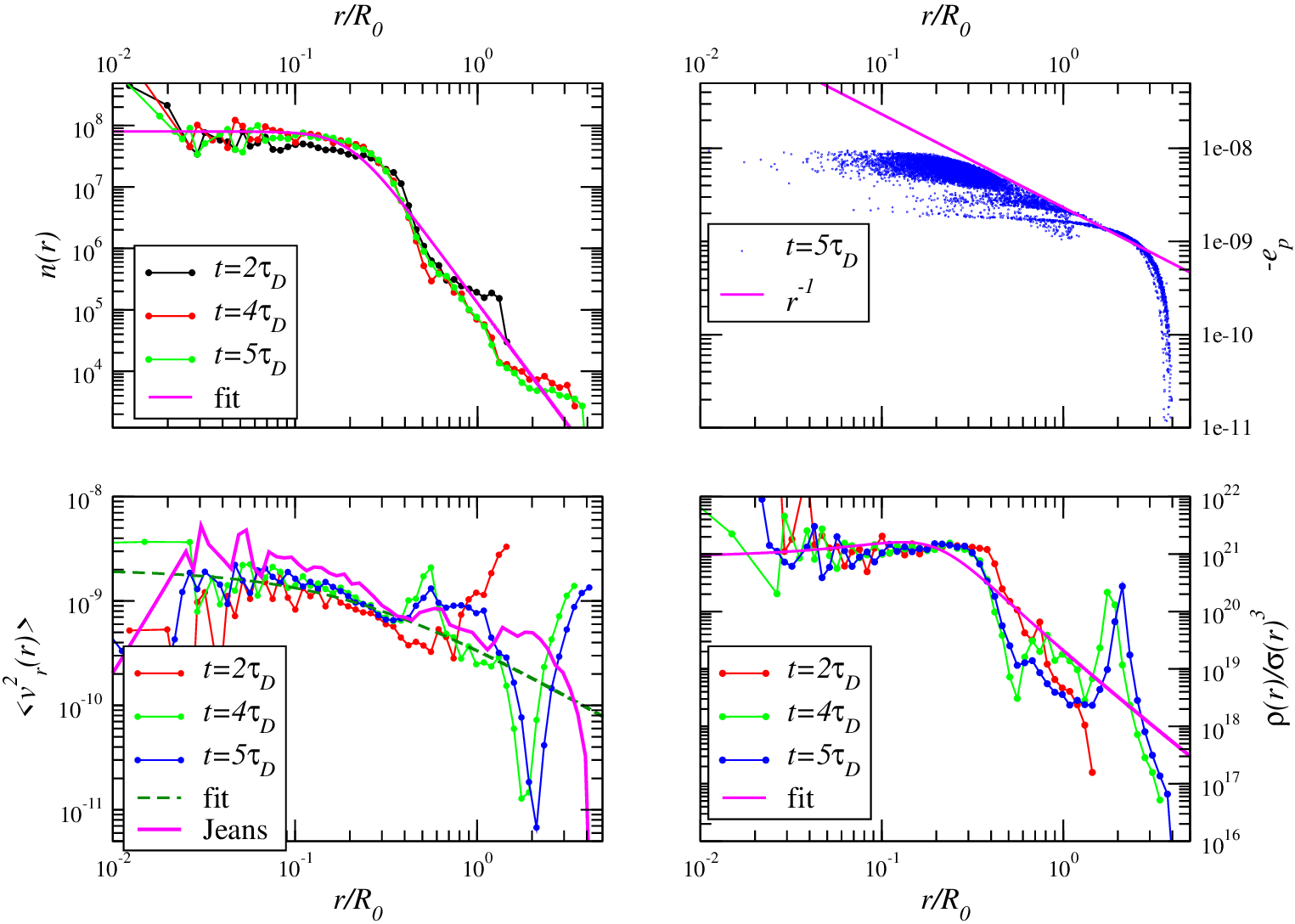}}
\par\centering
}
\caption{
As Fig.\ref{Fig1_b03} but for the case $b_0=-0.3$. }
\label{Fig1_b03}
\end{figure}

\begin{figure}
\vspace{1cm}
{
\par\centering \resizebox*{9cm}{8cm}{\includegraphics*{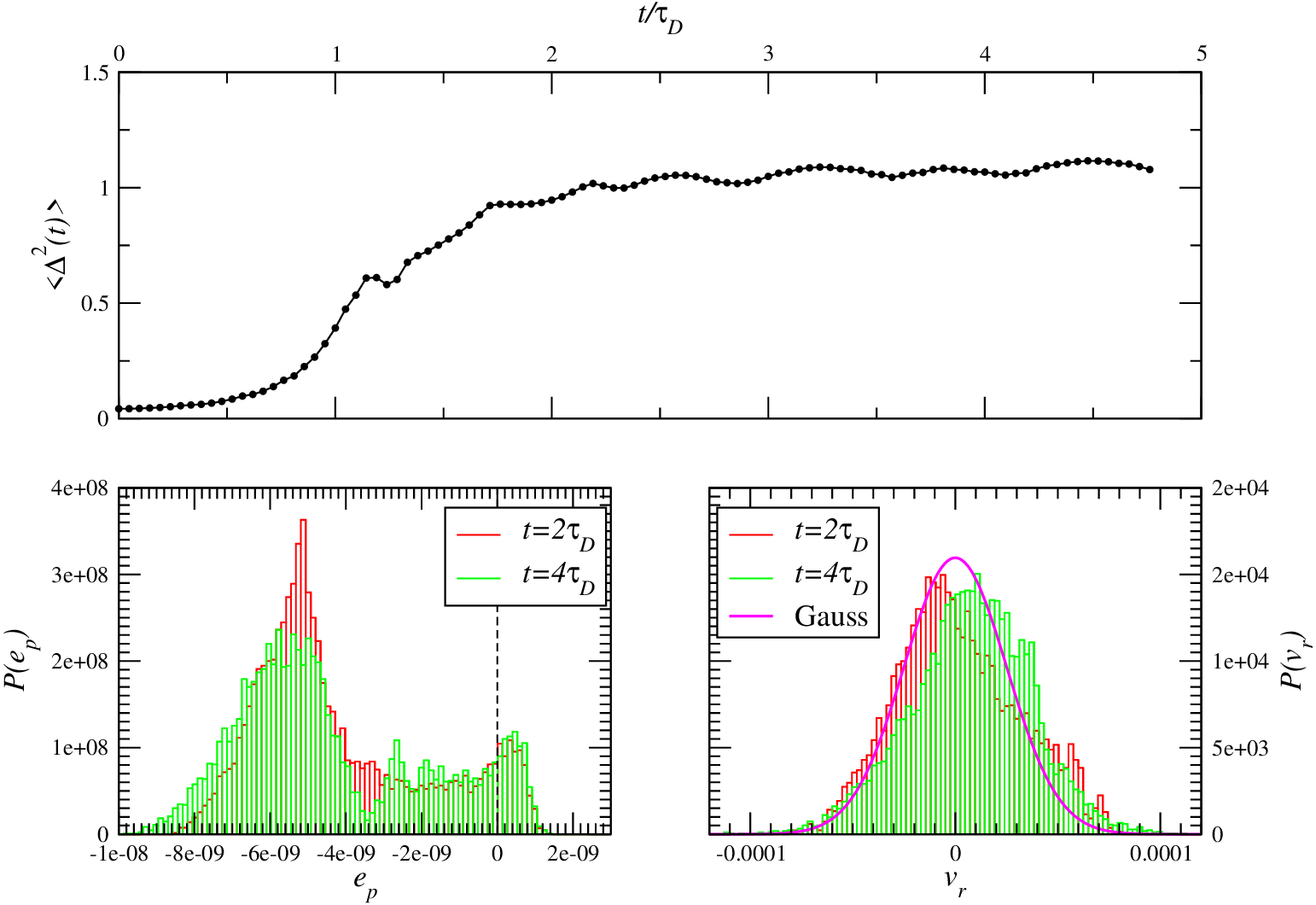}}
\par\centering
}
\caption{ 
As Fig.\ref{Fig2_b0} but for the case $b_0=-0.3$. }
\label{Fig2_b03}
\end{figure}

Behaviors similar to the ones shown in
Figs.\ref{Fig1_b0}-\ref{Fig2_b0} are found for the case in which the
initial virial ratio is $b_0=-0.3$ (see
Figs.\ref{Fig1_b03}-\ref{Fig2_b03}).  However, due a the non zero
initial velocity dispersion, the collapse is less peaked in time.  The
density profile is again well approximated by Eq.\ref{dp_b0}, but in
this case $r_c=0.2R_0$ i.e. it is about ten times larger than for the
$b_0=0$ case. Correspondingly we find $n_c(b_0=-0.3) < n_c(b_0=0)$,
i.e. the structure of the QSS is much less compact.  Also the
behaviors of $\langle v_r^2 (r) \rangle$ and of $\rho(r)/\sigma^3(r)$
are well described by Eqs.\ref{fitv2r}-\ref{fitpsd} although with
different parameters.  Finally $\Delta^2(t)$ shows that there is a
smaller exchange of energy during the collapse phase than in the
$b_0=0$ case, but still much larger than for $b_0<-1/2$.  In brief, in
this case the collapse is less violent and the fraction of particles
with positive energy for $t > \tau_D$ is greatly reduced with respect
to the $b_0=0$ case (see Fig.\ref{R0t}).


\section{The origin of the power law tail in the violent relaxation case} 
\label{model} 
 
We now introduce a simple physical model with the aim of describing
the dynamics of bound particles with $r>r_c$ when $b_0>b_0^c \approx
-1/2$.  We then show that this model allows us to capture the
essential ingredients that originate the power law $r^{-4}$ tail in
the density profile (Eq.\ref{dp_b0}).

\subsection{A simple physical model} 

 We suppose that the bound particles with $r>r_c$ for
$t>\tau_D$, are only subjected to the gravitational field of the core
with mass 
\be
\label{masscore}
M_c = 4 \pi \int_0^{r_c} \frac{n_c m}{1+(r/r_c)^4} r^2 dr 
\approx \frac{4 \pi}{3} n_c r_c^3 m \;,
\ee
so that the equation of motion for one of these particles 
is simply
\be
\label{eq_motion}
\frac{d^2r} {dt^2}  =  - \frac{G M_c}{r^2}  \;.
\ee
We can integrate Eq.\ref{eq_motion} to get 
\be
\label{eq_energy}
\frac{1}{2} \left(\frac{dr}{dt}\right)^2 = \frac{GM_c}{r} -\epsilon_0
\ee where we defined \be \epsilon_0 = \frac{GM_c}{r_0} -
\frac{1}{2}v_0^2 \;,
\ee 
and $r_0,v_0$ are respectively the initial position and velocity at
the initial time $t_0$.

The initial conditions at $t_0$ are specified as follows. We take the
origin of the time at $t_0=\tau_D$, i.e. the time of maximum collapse
of the system. In this situation particles are confined in a spherical
volume of radius $r_{min} \approx r_c$, the minimal radius of the
system during the collapse.  As particles forming the density
power-law tail must be bound, their energy is negative,
i.e. $\epsilon_0 = - e_p(\tau_D) > 0$ at $t=\tau_D$. We make the
hypothesis that this energy is conserved at later times.  This
hypothesis is both confirmed by the simulation (see below) and
compatible with the fact that the system is a quasi-stationary
equilibrium for $t>\tau_D$. Indeed, in a QSS --- defined in the mean
field limit --- each particle moves in a time independent potential,
and therefore has exactly fixed energy.  In principle, any change of
energy is due to finite $N$ effects, which are, however, relevant only
on a much long time scale than the one considered here.  { 
  Particles velocities are assumed to be oriented outwards, an
  hypothesis that agrees with the fact that after the collapse the
  velocity dispersion is by dominated by its radial component (see
  Fig.\ref{Fig_VelAn}).}

Bound particles may have a
maximum velocity such that $e_p^M=0$, i.e. for
\be
\label{v0M} 
v_0^M = \sqrt{ \frac{2GM_c}{r_0}} \;,
\ee 
so that also the velocity is bounded in $0< v_0 \le v_0^M$.
By defining 
\be
\label{rM}
r_H= \frac{GM_c}{2 \epsilon_0}
\ee
we can rewrite Eq.\ref{eq_energy}  as
\be
\label{eq_energy2}
\frac{dr}{\sqrt{2 r_H r -r^2}} = \sqrt{2 \epsilon_0} \frac{dt}{r}
\equiv d\eta \;, \ee 
where the last equality defines the variable
$\eta$.  Eq.\ref{eq_energy2} has solution in a parametric form \bea
\label{eq_energy3a}
&&
r(\beta) =  r_H (1 + \sin (\beta))
\\ 
\label{eq_energy3b}
&&
t(\beta) = \frac{r_H}{\sqrt{2 \epsilon_0}}  
\left(  \beta   - \cos(\beta) - \beta_0   +  \cos(\beta_0) \right)
\eea
where 
\be
\label{eq_energy4}
\beta_0= \sin^{-1}\left( \frac{2 r_0}{r_H} -1 \right) \;. 
\ee
Thus for $\beta=\beta_0$ we have $t=0$ and $r(0)=r_0$.  
Therefore we obtain 
\be
\label{rv0} 
r(\beta) \le r = \frac{2GM_c} {(v_0^M)^2-(v_0)^2} \;\; \forall \beta \;,
\ee
where the equality holds for the peaks of the sinusoidal function. 
By inverting Eq.\ref{rv0} and using Eq.\ref{v0M} we find
\be
\label{v0r}
v_0 \approx 
\sqrt{ (v_0^M)^2 - \frac{2GM_c}{r} } = v_0^M \sqrt{1- \frac{r_0}{r}} \;.
\ee

\begin{figure}
\vspace{1cm}
{
\par\centering \resizebox*{9cm}{8cm}{\includegraphics*{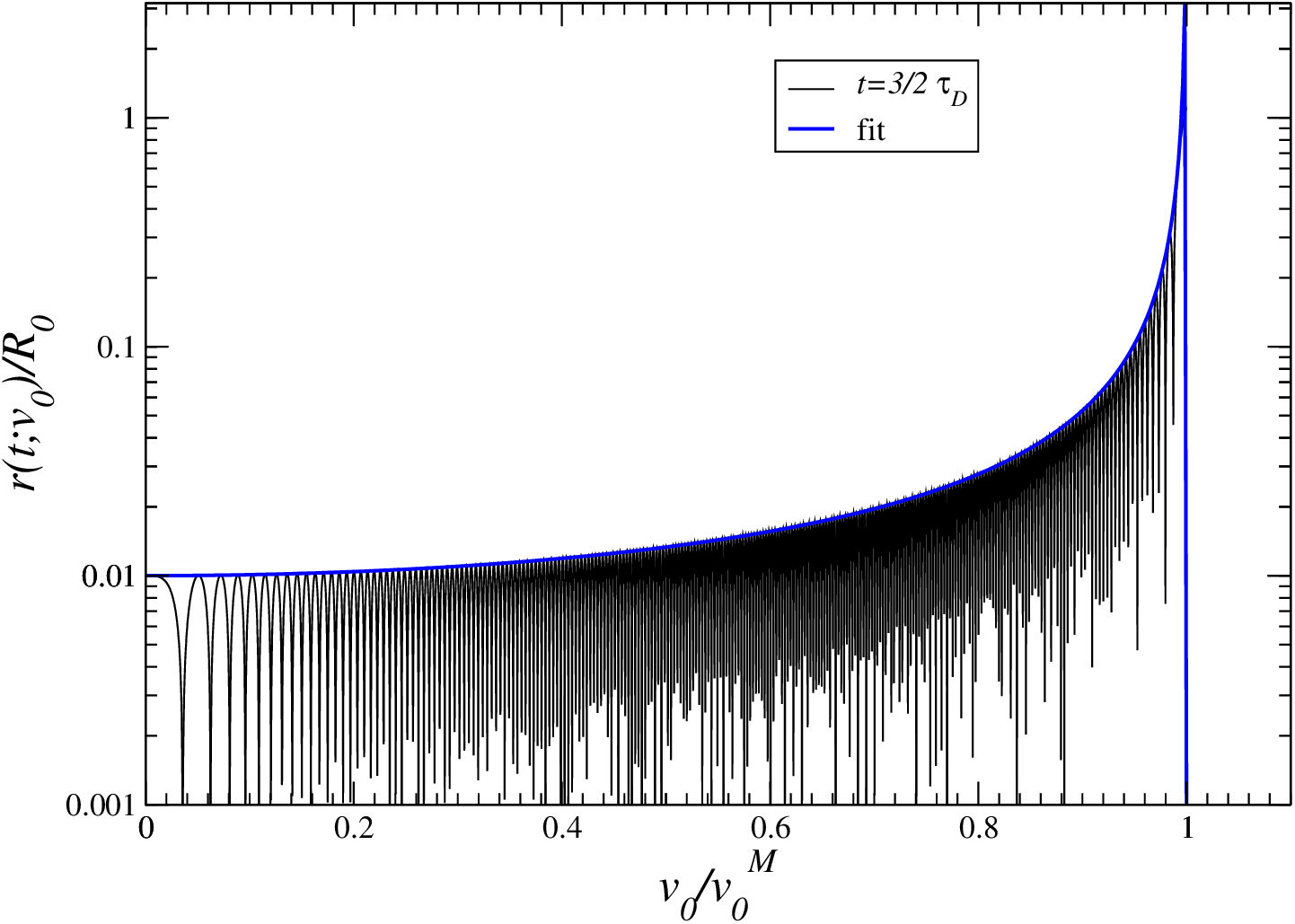}}
\par\centering
}
\caption{Behavior of the distance $r(t;v_0)$ reached by a particle after a
  time $1/2 \tau_D$ from the collapse, as a function of its initial
  velocity $v_0 \in[0,v_0^M]$ (from Eqs.\ref{eq_energy3a}-\ref{eq_energy3b}).  
  The
  behavior of Eq.\ref{rv0} is also plotted.
\label{r_vo}}
\end{figure}

\begin{figure}
\vspace{1cm} { \par\centering
  \resizebox*{9cm}{8cm}{\includegraphics*{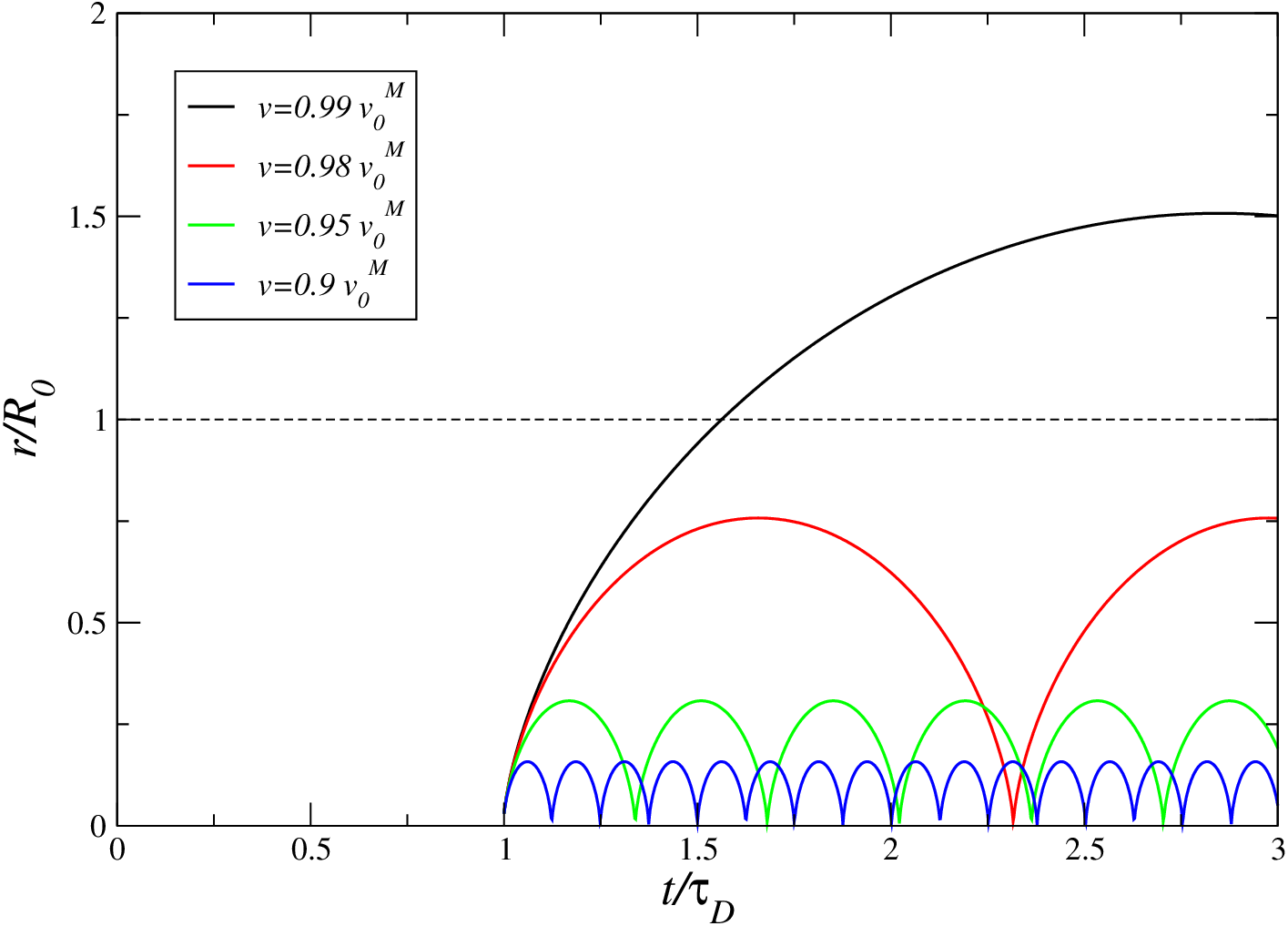}} \par\centering }
\caption{Behavior of the distance $r(t)$ as a function of time for
  different values of its initial velocity $v_0$ (from
  Eqs.\ref{eq_energy3a}-\ref{eq_energy3b}).
\label{r_time}}
\end{figure}

The behavior of the radial distance $r$ as a function of the initial
velocity $0 < v_0 \le v_0^M$, computed from
Eqs.\ref{eq_energy3a}-\ref{eq_energy3b}, is plotted in Fig.\ref{r_vo}
for $t = 3/2 \tau_D$.  
Similarly in Fig.\ref{r_time} it is plotted $r(t)$ for
different values of the initial velocity.
One may note that only the high velocity particles, for which $v_0
\approx v_0^M$, may reach a distance of the order of the initial
system's radius $R_0$. As time passes, the particles with the highest
velocity increases their radial distance to   $r > R_0$. 
At a time of the order of $\approx 3/2 \tau_D$ the structure
has already reached its (almost) asymptotic shape for $r_c < r \ltapprox
R_0$: while at larger scales and at later times, a few particles 
may arrive to larger and larger distances.

\subsection{Shape of the density profile} 

Let us now compute, under some simple approximations, the density
profile resulting from this simple physical model. We suppose that all
particles have, at the same initial time $\tau_D$, the same initial
position $r_0 \le r_c$.  In this approximation, given a certain
distribution of initial velocities $p(v_0)$, we find that the radial
density profile, for $t \gtapprox 2 \tau_D$ (see Fig.\ref{r_time}), is
given by
\be
\label{dentheo} 
n(r) \approx 
 \frac{1}{4 \pi r^2} N p(v_0) \frac{dv_0}{dr} = \frac{G M_c N p(v_0)}{4 
  \pi r^4 \sqrt{ (v_0^M)^2 - \frac{2GM_c}{r} } }  \;. 
\ee
One may note that from Eq.\ref{v0r} we find that $v_0 \approx v_0^M$
for $r >r_c$, so that in this limit Eq.\ref{dentheo}  becomes 
\be
\label{dentheo_approx} 
n(r) \approx  \frac{G M_c N}{4 \pi r^4 v_0^M}  p(v_0^M) \;,
\ee
thus showing the $r^{-4}$ decay in the best fit of measured $n(r)$
(Eq.\ref{dp_b0}).

Although in the derivation of Eq.\ref{dentheo} we made important
simplifications, we now show that the hypotheses used allow us to
capture the main elements of the problem. We may relax these
assumptions by allowing that the initial particle positions $r_0$ also
have a certain PDF $f(r_0)$.  In this case we need to integrate
Eqs.\ref{eq_energy3a}-\ref{eq_energy3b} numerically as follows:
\begin{itemize}

\item  We extract the initial conditions
$[r_0,v_0]$, such that $0< r \le r_c$ and $0 < v_0 \le v_0^M$, from the 
assigned initial position and velocity PDFs $f(r_0)$ and $p(v_0)$. 

\item We fix the time $t>\tau_D$ at which
we compute the profile.

\item  We then find from  Eq.\ref{eq_energy3b} the
value $\beta=\beta(t,r_0,v_0)$ and then from  Eq.\ref{eq_energy3a}
$r(t)=r(t,r_0,v_0)$. 

\item We repeat the iteration $N$ times and we can thus
construct numerically, from the resulting distribution of distances
$r(t)$, the density profile at time $t$.  
\end{itemize} 

\begin{figure}
\vspace{1cm}
{
\par\centering \resizebox*{9cm}{8cm}{\includegraphics*{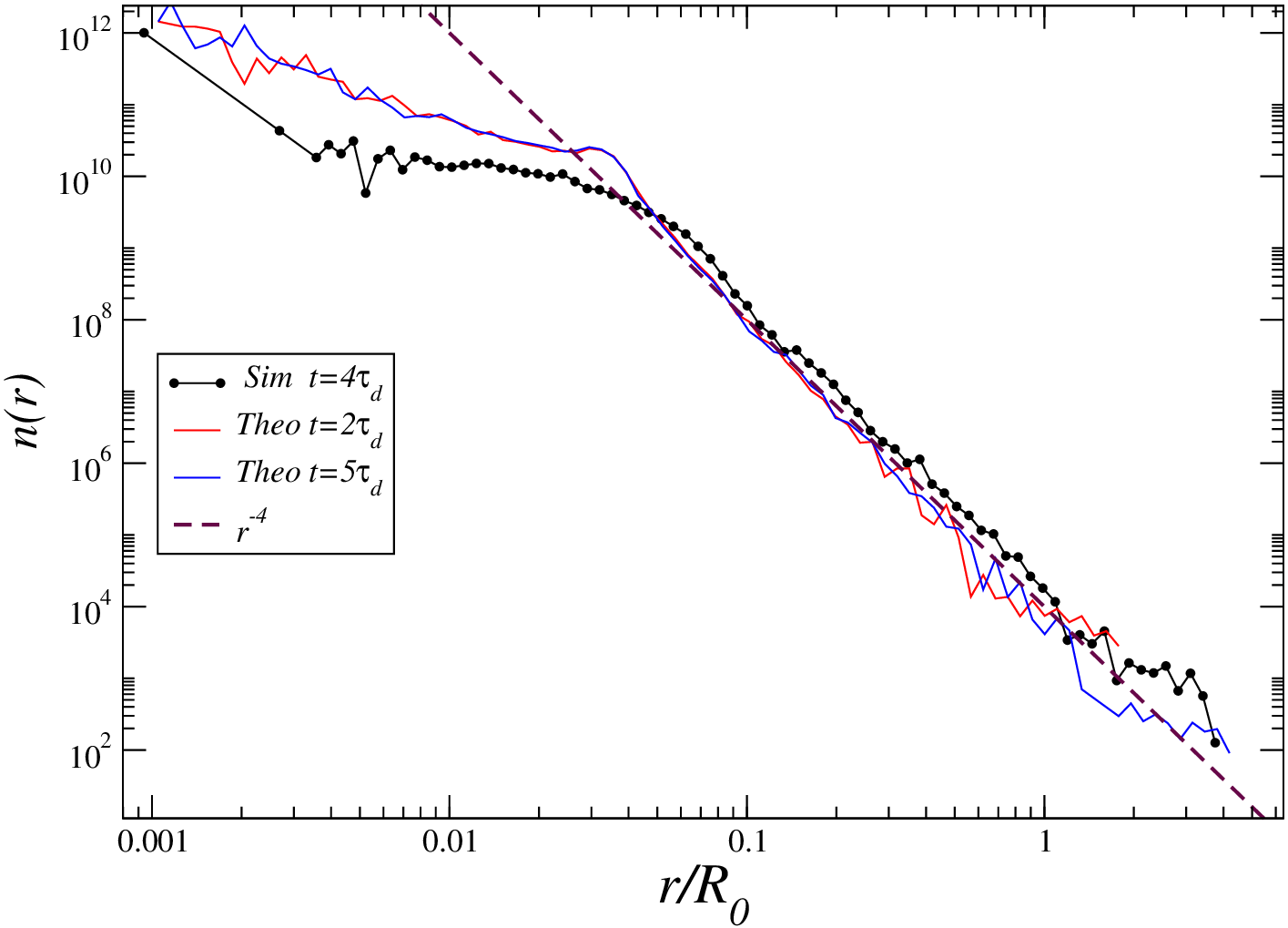}}
\par\centering
}
\caption{Behavior of the {  theoretical} 
radial density $n(r)$ computed with the
  Monte-Carlo method described in the text at different times.
  The normalization is arbitrary.} 
\label{theo1_OBC_inf_b0}
\end{figure}

As an example (see Fig.\ref{theo1_OBC_inf_b0}) we have assumed a
Gaussian velocity distribution $p(v_0)$ with zero mean and variance
$(v_0^M)^2/\alpha$, where $\alpha>1$ is a free parameter, and we have
considered only particles with $v_0 > 0$. In addition we have taken  
a uniform distance $r_0$ distribution such that 
$f(r_0)\ne 0$ only for $ r_0 \in[0, r_c]$.
We have tested that no sensible differences are detected as long as
$\alpha$ is not large enough to get a too small value of $p(v_0^M)$:
in particular, 
when $Np(v_0^M) <1$ a particle cannot neither be ejected nor form the
$r^{-4}$ tail.  In principle, one should also consider the
correlations between $r_0$ and $v_0$ and the fact that particles have
a distribution of initial times: however these complications
do not significantly  alter the result of Eq.\ref{dentheo_approx}. 

Finally we note that as time passes, for $t>\tau_D$, the power-law
tail extends to larger and larger scales (see
Fig.\ref{Fig1_b0}). This is simply explained by the behaviors shown in
Figs.\ref{r_vo}-\ref{r_time}: the largest distance reached by the
particles with highest velocity also increases with time, although the
precise  manner in which this occurs 
depends in a very detailed way on the value on the 
properties of $p(v_0)$ for $v_0 \rightarrow v_0^M$.


\section{The critical value of the initial virial ratio} 
\label{critical_value}

We now consider the question of what determines the
critical value of the initial virial ratio for having or not ejection
(and thus the formation of the $n(r) \sim r^{-4}$ power law tail) to
be $b_0^c \approx -1/2$. We would like to stress that  the
precise value $b_0^c$ must be a function of $N$,
as  collisional and discrete effects, although represent
perturbations, are also present in the collapse phase
 \citep{paper1}.

As mentioned in Sect.\ref{ejection}, the mechanism of {  energy and
  mass} ejection is based on the fact that a fraction of the
particles, and particularly those that lie at the boundary of the
system at the initial time, lag behind with respect to the others
during the collapse, i.e. at $t < \tau_D$.  Particles in the bulk
collapse approximately satisfying the condition of homologous
contraction (Eq.\ref{scm1}). The question is whether this is also
satisfied when $b_0 <0$.

\begin{figure}
\vspace{1cm} { \par\centering
  \resizebox*{9cm}{8cm}{\includegraphics*{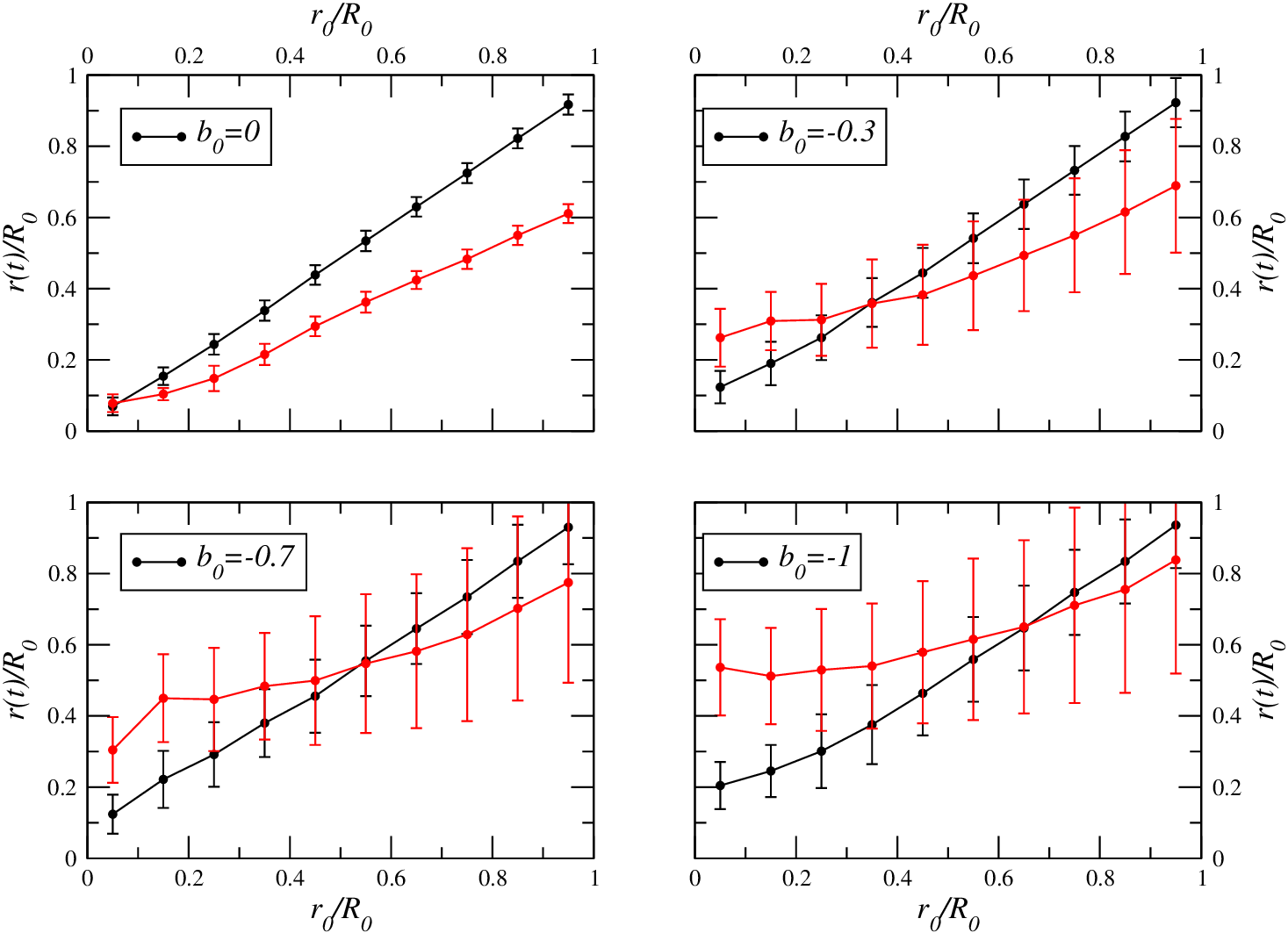} }
  \par\centering }
\caption{Behavior of the position at time $t$ (black dots $t=1/4
  \tau_D$, red dots $t=3/4 \tau_D$) as a function of the initial
  position $r_0$, averaged in shells, for different values of the
  initial virial ratio (see caption).
\label{r0rt} }
\end{figure}

In Fig.\ref{r0rt} we plot, for different values of the initial virial
ratio $b_0$, particle positions at time $t$ (for $t=1/4 \tau_D, 3/4
\tau_D$) as a function of their positions at time $t=0$. We have
considered an average in shells where these are taken at $t=0$; we
then plot the average value in each shell together with the
r.m.s. error. One may see that for $b_0=0$ the two curves do not
overlap, while this marginally occurs for $b_0=-0.3$.  In this case
the homologous contraction is a reasonable approximation of the
collapse.

For smaller values of the initial virial ratio, i.e. $b_0=-0.7, -1$,
there is a substantial overlapping of the curves at different times,
which means that particles originally belonging to different shells
interchange their positions.  This implies that the collapse cannot be
anymore approximated as homologous because different shells largely
overlap well before $\tau_D$.
The key mechanism of the growth of the time lag is thus eliminated
when particles have initially high enough velocity dispersion, as 
different shells cross each other well before $\tau_D$. Thus particles
from the outer shell arrive at different time at the center and they
do not gain the necessary energy to escape from the system.

{  Finally it should be noted that the analysis presented in this
  section holds only for water bag initial conditions, and that a
  different initial spatial and velocity distribution can lead to very
  different behaviors.  For instance, \citet{trenti2005} have found
  that initial conditions with equal virial ratio but different
  spatial distributions, as for example by generating clumpy
  distributions, lead to significantly less mass ejection.  A similar
  result was found by \cite{trentibertin2006} when particles were
  initially distributed in a shell rather than in a sphere.  Given
  that the precise amount of mass and energy ejection is determined by
  the coupling of the growth of perturbation with the finite size of
  the system it is difficult to develop a general argument which is
  valid for different initial conditions and a more detailed
  consideration of each case is needed.}


\section{Discussion and conclusion}
\label{conclusion}

The collapse of an isolated, uniform and spherical cloud of massive
particles interacting by Newtonian gravity represents a paradigmatic
example for the formation of quasi equilibrium states. It is indeed
well known, since the earliest N-body simulations that, when initial
velocities are set to zero, this system collapses in a relatively
short time scale $\tau_D \approx \sqrt {G \rho_0}^{-1}$ reaching a
configuration which satisfies the virial theorem $b(t) = 2K/W
\approx -1$.  The collective relaxation process acting on such a short
time scale and the statistical properties of the formed quasi
equilibrium state have been considered in this paper both by
performing numerical simulations and by an attempt to elaborate a
physical model able to capture the essential elements of the problem.

In particular, the initial conditions of simulations are generated by
randomly placing $N$ particles with average mass density $\rho_0$ in a
spherical volume and characterized by different initial virial ratio
$0 \ge b_0 \ge -1$. Thus, only two parameters, {  i.e.,} $N, b_0$, define the
initial conditions: in this paper we have varied $b_0$ in the range
$[-1,0]$ while in \cite{paper1} we have considered, for the case
$b_0=0$, simulations with different number of points $N$.

The system thus collapses under its self-gravity and then it forms a
virialized structure.  This does not represent an equilibrium state in
the thermodynamics sense. Indeed, two body collisions, which have a
time scale of about $\tau_2 \approx \tau_D \ln(N)/N$ will cause a slow
evaporation of particles from it \citep{binney}. For this reason the
state formed after $\tau_D$ is called a quasi stationary state (QSS)
\citep{Dux03,Campa08} .

By considering N-body simulations with different initial virial
ratios, we have identified a critical initial virial ratio $b_0^c
\approx -1/2$ separating the formation of two qualitatively different
kind of QSS. When $-1\le b_0<-1/2$ the collapse consists in a series
of damped oscillations, {  the first of which} one has larger
amplitude.  The system thus approximately maintains its original
size. The density profile characterizing the virialized QSS is well
fitted by the predictions of the Lynden-Bell {  distribution with a
  cut-off by considering the system confined in a box}
\citep{levin_etal_2008}. {  This is } characterized by an abrupt
decay of the density at a scale $r_c \approx R_0$.  The Lynden-Bell
predictions strictly holds for a confined system: however {  it was
  found that} the theory does not depend sensibly on the cut-off
value.  This approach is thus useful to understand the properties of
gentle kind of collective relaxation which occurs when the systems is
initially in a configuration which is close enough to the virial
equilibrium.

On the other hand, for $-1/2 \le b_0 \le 0$ the system size undergoes
to a large compression and a part of its mass and energy is ejected.
Finite $N$ fluctuations in the initial spatial particle configuration
generate density perturbations which grow during the collapse. When
$b_0=0$ such a dynamical problem can be treated, when boundaries
effects are neglected (i.e. in the limit $R_0 \rightarrow \infty$) as
the growth of perturbations in a contracting universe.  When
fluctuations at a scale $r_{min}$ of order of the size of the system
go non-linear the collapse is stopped
\citep{aarseth_etal_1988,boily+athanassoula_2006,boily_etal_2002,paper1}.

During such the collapse some particles gain enough kinetic energy
that can be ejected from the system. The ejection mechanism was
studied in detail by \cite{paper1} where it was shown to be related to
a boundary effect.  Particles initially placed close to the boundaries
arrive later than the others toward the center, moving, for a short
time interval, in a rapidly varying potential field generated by the
particles which have already inverted their motion from inwards to
outwards. In this way they gain some kinetic energy, so that some
particles have positive energy $e_p>0$.  The density profile $n(r)$ of
the bound system formed after the collapse is characterized by a core,
where $\rho(r) \sim const.$ and by an halo in which $n(r) \sim
r^{-4}$. The former behavior can be understood by considering that the
distribution function of the core is given by that of a fully
degenerate Fermi gas: this can be obtained again from the {  cut-off
  Lynden-Bell distribution}, by letting the cut-off to extend to
infinity so that particles can move far away from the center. In this
case it forms a core-halo structures, with a dense core a diluted halo
\citep{levin_etal_2008}.  The Lynden-Bell theory cannot however be
used to {  derive} the properties of the halo, i.e.  that the radial
density decays as $n(r) \sim r^{-4}$.
 
In order to understand the formation of such a power-law tail we have
introduced a simple physical model based on a few ingredients, namely
that that: (i) at the time of maximum contraction, i.e. $t \approx
\tau_D$, particles are confined in a small phase-space region, (ii)
particles energy may be close to, or larger than, the escape one and
(iii) particles forming the power-law tail move in a central and
constant gravitational potential generated by the mass of the core
$M_c$ at $r<r_c$. With these assumptions a density profile with a
power-law tail is naturally formed.  We conclude that the behavior
$n(r) \sim r^{-4}$ is the typical density profile that is obtained
when the initial conditions are cold enough that ejection {  of mass
  and energy} occurs.

The critical virial ratio $b_0^c$ separating the two situations in
which the power-law profile is formed, and mass ejection occurs, can
be understood by considering that when the initial velocity dispersion
is large enough the contraction is no more homologous.  Therefore
different shells may overlap before the final collapse phase at
$t\approx \tau_D$ and the mechanism underlying the gain of energy for
the outer particles cannot be working anymore.

Finally it is interesting to note that cold dark matter halos in
cosmological simulations
\citep{navarro1,Moore1,Moore2,navarro2,Hansen,navarro,merritt2006}
display a density profile such that $n(r) \sim r^{-1}$ at small scales
and $n(r) \sim r^{-3}$ at large scales: these behaviors are not
observed to form from the simple initial conditions we have chosen
\footnote{  Although \cite{graham2006} found a steeper slope at
  large scales.}  Also the phase space density has a different shape,
decaying as $\rho/\sigma^3 \sim r^{-1.875}$ at all scales in
cosmological simulations \citep{navarro} while it displays a
$r^{-5/2}$ behavior only at large enough scales, i.e. $r>r_c$, in the
case of structures formed from the initial conditions we considered
(when $b_0^c >-1/2$).  This difference maybe originated by that the
fact that cosmological halos are formed from more complicated { 
  initial conditions than the case we considered}.  However, one
should also consider that cosmological halos are formed in a complex
backgrounds so that the hypothesis that they are isolated structures
maybe not be {  a valid assumption}. In addition, there is a
continuous mass accretion so that neither the total mass nor the total
energy are conserved.  A more focused study of these features will be
presented in a forthcoming work.
\bigskip


I acknowledge Roberto Capuzzo-Dolcetta, Massimo Cencini, Umberto
Esposito, Andrea Gabrielli, Michael Joyce, Yan Levin and Tirawut
Worrakitpoonpon for useful discussions and comments.



\bibliographystyle{mn2e}

\end{document}